\documentclass{frontiersSCNS} 

\usepackage{url}

\usepackage{graphicx}
\usepackage{txfonts}
\usepackage{amssymb}
\usepackage{enumerate}
\newcommand{\BE}{\begin{equation}}
\newcommand{\EE}{\end{equation}}
\newcommand{\BA}{\begin{eqnarray}}
\newcommand{\EA}{\end{eqnarray}}

\renewcommand{\vec}[1]{{\boldsymbol #1}}

\newcommand{\bb}{\vec B}

\newcommand{\xx}{ \vec x}

\newcommand{\kevr}[1]{{\bf \color{magenta}{[]}} \color{black}}  

\newcommand{\dougr}[1]{{\bf \color{red}{[]}} \color{black}}

\newcommand{\sarahr}[1]{{\bf \color{blue}{[]}} \color{black}} 

\newcommand{\eq}[1]{Equation~(\ref{eq:#1})} 
 
\newcommand{\eqss}[2]{Equations~(\ref{eq:#1}) -- (\ref{eq:#2})} 
\newcommand{\sect}[1]{Section~\ref{sec:#1}}

\newcommand{\app}[1]{Appendix~\ref{app:#1}}

\newcommand{\tab}[1]{Table~\ref{tab:#1}}

\newcommand{\fig}[1]{Figure~\ref{fig:#1}}

\newcommand{\eg}{\textit{e.g.}, }
\newcommand{\ie}{\textit{i.e.}, }

\def\keyFont{\fontsize{8}{11}\helveticabold }
\def\firstAuthorLast{Dalmasse {et~al.}} 
\def\Authors{K. Dalmasse\,$^{1,*}$, D. W. Nychka\,$^{2}$, S. E. Gibson\,$^3$, Y. Fan\,$^3$ and N. Flyer\,$^2$}
\def\Address{
$^{1}$CISL/HAO, National Center for Atmospheric Research, Boulder, CO, USA \\
$^{2}$CISL, National Center for Atmospheric Research, Boulder, CO, USA \\
$^{3}$HAO, National Center for Atmospheric Research, Boulder, CO, USA
}
\def\corrAuthor{K\'evin Dalmasse}
\def\corrAddress{CISL/HAO, National Center for Atmospheric Research, P.O. Box 3000, Boulder, CO 80307-3000, USA}
\def\corrEmail{dalmasse@ucar.edu}

\begin{document}
\onecolumn
\firstpage{1}

\title[Method for diagnosing the 3D coronal magnetic field]{{ROAM: a Radial-basis-function Optimization Approximation Method} for diagnosing the three-dimensional coronal magnetic field}
\author[\firstAuthorLast ]{\Authors}
\address{}
\correspondance{}
\extraAuth{}
\topic{Coronal Magnetometry}

\maketitle


\begin{abstract}

\section{
The Coronal Multichannel Polarimeter (CoMP) routinely performs coronal polarimetric 
measurements using the Fe XIII 10747 $\AA$ and 10798 $\AA$ lines, which are 
sensitive to the coronal magnetic field. However, inverting such polarimetric measurements 
into magnetic field data is a difficult task because the corona is optically thin at these 
wavelengths and the observed signal is therefore the integrated emission of all the plasma 
along the line of sight. To overcome this difficulty, we take on a new approach that combines 
a parameterized 3D magnetic field model with forward modeling of the polarization signal. 
For that purpose, we develop a new, fast and efficient, optimization method for model-data 
fitting: the Radial-basis-functions Optimization Approximation Method (ROAM). Model-data 
fitting is achieved by optimizing a user-specified log-likelihood {function} that quantifies 
the differences between the observed polarization signal and its synthetic/predicted analogue. 
Speed and efficiency are obtained by combining sparse evaluation of the magnetic model 
with radial-basis-function (RBF) decomposition of the log-likelihood {function}. The RBF decomposition 
provides an analytical expression for the log-likelihood {function} that is used to inexpensively estimate 
the set of parameter values optimizing it. We test and validate ROAM 
on a synthetic test bed of a coronal magnetic flux rope and show that it performs well 
with a significantly sparse sample of the parameter space. We conclude that our optimization 
method is well-suited for fast and efficient model-data fitting and can be exploited for converting 
coronal polarimetric measurements, such as the ones provided by CoMP, into coronal magnetic 
field data.
}
\tiny
 \keyFont{ \section{Keywords:} Sun: corona, Sun: magnetic fields, Sun: infrared, Methods: statistical, Methods: radial basis functions} 
\end{abstract}
   
%

\section{Introduction} \label{sec:S-Introduction}

%
%
Modification to the polarization of light is one of the many signatures of a non-zero 
magnetic field in the solar corona, and more generally, in the solar atmosphere 
\citep[\eg][and references therein]{Stenflo15}. Several mechanisms producing 
or modifying the polarization of light have been observed and studied in the solar 
corona at different wavelengths including, but not limited to, the Zeeman and Hanle effects 
\citep[see \eg][and references therein]{Hale08,Hanle24,Bird85,White97,Casini99,Lin04,Gibson16}. 
The former induces a frequency-modulated polarization while the latter induces a depolarization 
of scattered light \citep[\eg][]{SahalBrechot77,Bommier82,Rachmeler13,LopezAriste15}. 
Both mechanisms allow us to probe the strength and direction of the coronal magnetic field. 
Coronal polarization associated with these two mechanisms is currently measured above 
the solar limb by the Coronal Multichannel Polarimeter from forbidden coronal lines such 
as the Fe XIII lines \citep[10747 $\AA$ and 10798 $\AA$;][]{Tomczyk08}. For these two lines, 
the circular polarization signal is dominated by the Zeeman effect while the linear polarization 
signal is dominated by the Hanle effect \citep[\eg][]{Judge06}.

%
%
Translating the polarization maps of CoMP into magnetic field maps is a challenging 
task. The main difficulty is that the solar corona is optically thin at these wavelengths 
\citep[\eg][]{Rachmeler12,Plowman14}. The observed signal is therefore the integrated 
emission of all the plasma along the 
{line of sight (LOS).} 
Hence, the polarization maps cannot, 
in general, be directly inverted into 2D maps of the plane-of-sky (POS) magnetic field. 
On the other hand, extracting individual magnetic information at specific positions along the LOS 
is extremely difficult without stereoscopic observations \citep[\eg][]{Kramar14}. Another limitation 
is that the Hanle effect associated with the aforementioned forbidden infrared lines operates 
in the saturated regime \citep[\eg][]{Casini99,Tomczyk08}. Accordingly the linear polarization 
signal measured by CoMP is sensitive to the direction of the magnetic field but not its strength. 
Deriving the magnetic field associated with the polarization maps of CoMP therefore requires 
a different approach than the single point inversion that can be done with, \eg photospheric 
polarimetric measurements.

%
%
The alternate approach we propose to follow is to combine a parameterized 3D 
magnetic field model with forward modeling of the polarization signal observed 
by CoMP. For that purpose, we take advantage of the Coronal Line Emission 
(CLE) polarimetry code developed by \cite{Casini99} and integrated into 
the FORWARD package. FORWARD\footnote{\url{http://www.hao.ucar.edu/FORWARD/}} 
is a Solar Soft\footnote{\url{http://www.lmsal.com/solarsoft/}} IDL package designed 
to perform forward modeling of various observables including, \eg visible/IR/UV 
polarimetry, EUV/X-ray/radio imaging, and white-light coronagraphic observations 
\citep{Gibson16}. The goal is then to optimize a user-specified likelihood {function} comparing 
the polarization signal predicted by FORWARD to the real one and find the parameters 
of the magnetic field model such that the predicted signal fits the real data.

%
%
In the present paper, we develop and test a new method for performing fast and 
efficient optimization in a $d$-dimensional parameter space that may be used 
for converting the polarization observations of CoMP into magnetic 
field data. The optimization method, called ROAM (Radial-basis-functions Optimization 
Approximation Method) is designed to be general enough so that it can be applied 
independently of the dimension and size of the parameter space, the 3D magnetic 
field model, the type of observables (provided that one can forward model them), 
and the form of the likelihood {function} used for comparing the predicted signal to the real one. 
ROAM is introduced in \sect{S-Method}. \sect{S-Results} describes the results 
of multiple applications of ROAM to a synthetic test bed as validation of the optimization 
method. Our conclusions are then summarized in \sect{S-Conclusions}.


\section{Method} \label{sec:S-Method}

The goal of this paper is to propose a model-data fitting method to be used for {\it near-real-time} 
3D reconstruction of the solar coronal magnetic field. This requires developing a fast 
and efficient method for searching for the set of values of the model parameters that optimize 
a pre-defined function quantifying the differences between the predicted (or forward-modeled) 
and real data. 
{Although similar approaches are standard in engineering \citep[\eg][]{Jones98}, we propose 
a simplified version and tailored to the context of solar physics.} 
The proposed method, ROAM, combines {the computation of a log-likelihood function} 
on a sparse sample of the parameter space with function approximation and is based 
on the five following steps:
\begin{enumerate}
	\item Sparse sampling of the parameter space is performed using {\it Latin Hypercube Sampling} 
		\citep[LHS;][]{McKay79,Iman81}. LHS is a statistical method for generating a random 
		sample of the parameter values in a $d$-dimensional space. For a $d$-dimensional space of $n^d$ 
		points ($n$ is the number of points for each dimension), LHS creates a set, $\{ \xx_i \}$, of $n$ 
		independent points or $d$-vectors of the parameter space (an example is given \fig{Fig-LHC-examples}) 
		that will be referred to as the {\it design} in the following.

  \begin{figure}
   \centerline{\includegraphics[width=0.75\textwidth,clip=]{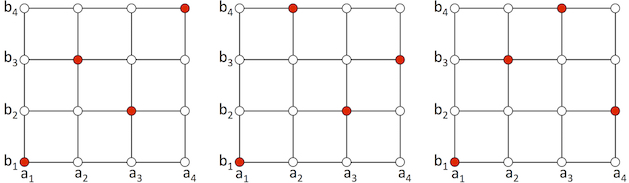}
              }
              \textbf{\refstepcounter{figure}\label{fig:Fig-LHC-examples} Figure \arabic{figure}.}{ Example of three {\it designs} generated via {\it latin hypercube sampling} (LHS) in 2D (red points). Note that each of these designs only possesses one point per column and per row, which is a special feature of LHS.}
   \end{figure} 

	\item The model is computed for each point, $\xx_i$, of the design and used to generate the corresponding 
		predicted observation, $y \left( \xx_i \right)$, to be compared with the ground truth, $y_{gt}$ 
		(which is either an actual observation or a synthetic one for test beds using analytical models 
		or numerical simulations).
	\item The set of predicted observations, $\{ y (\xx_i) \}$, is then compared to the ground truth by means 
		of a user-specified log-likelihood function
		\BE	\label{eq:Eq-General-Likelihood}
			\ell (\xx_i) = \log \mathcal{L} = f \left( y (\xx_i) - y_{gt} \right) \,,
		\EE
		where $\mathcal{L}$ is the likelihood {function}, $\xx_i$ is a $d$-vector of the design and $f$ is a general, 
		user-specified, well-behaved, scalar function. 
		{Typically, the likelihood function simplifies to depend on the difference 
		between the observations and the predicted values and function $f$ reflects that.} 
		An explicit expression 
		of $f$ is given in the section of each test considered in this paper (see \sect{S-Results}).
	\item This log-likelihood {function} is then approximated using {radial-basis-function} (RBF) decomposition 
		\citep[see \eg][]{Powell77,Broomhead88,Buhmann03,Nychka15}
		\BA	\label{eq:Eq-RBF-Likelihood}
			\ell (\xx) \approx \hat{\ell} \left( \xx \right) & = & \sum^{n}_{j=1} a_j \varphi_j \left( \Vert \xx - \xx_j \Vert \right) + \sum^{\binom{p+d}{p}}_{j=1} b_j  \psi_{j} \left( \xx \right)    \,, \\
			\label{eq:Eq-RBF-1}
			\varphi_j ( \Vert \xx - \xx_j \Vert ) & = & \Vert \xx - \xx_j \Vert^{2m-d} \log \left( \Vert \xx - \xx_j \Vert \right) \,,  \textrm{if $d$ is even}      \,, \\
			\label{eq:Eq-RBF-2}
			 & = & \Vert \xx - \xx_j \Vert^{2m-d}  \,,  \textrm{if $d$ is odd}    \,,
		\EA
		where $\varphi_j$ is the $j$-th {RBF} centered at point $\xx_j$ of the design, 
		$\Vert \cdot \Vert$ is the usual Euclidean norm, $m \in \mathbb{N}$ is such that $2m-d > 1$, 
		and $\{ \psi_{j} \}$ is a set of polynomials up to degree $p$ in the dimension $d$ of the problem 
		{with the constraint $p \le m - 1$. In the following, we always use $p = m-1$}. 
		When periodic components of the $d$-space exist, the value of $d$ must be modified for the RBF 
		decomposition to take the periodicities into account (an example and further details on handling 
		periodic components are provided in \app{A-Periodic-Variables}). Note that the particular choice 
		of RBFs, {$\varphi_j$}, in \eqss{Eq-RBF-1}{Eq-RBF-2} is called a {\it Polyharmonic Spline} 
		\citep[see \eg][]{Duchon77,Madych90} and that 
		{the polynomial term in \eq{Eq-RBF-Likelihood} is not 
		a regularization term but an additional term that directly comes from the definition of Polyharmonic 
		Splines as minimizers of the energy functional 
		$\int_{\mathcal{V} \subset \mathbb{R}^d} | \nabla^m g|^{2} \mathrm{d} \xx$ 
		(which is not modified by adding polynomials of order $p \le m-1$ to $g$). Although required 
		from the definition of Polyharmonic Splines, this polynomial term is particularly beneficial 
		for improving the fitting accuracy and extrapolation away from the RBF centers $\xx_j$, 
		while also ensuring polynomial reproductibility.} 
		Note also that the $a_j$ and $b_j$ are coefficients determined 
		from the set of $n$ equations provided by the constraint (the detailed derivation of the coefficients 
		is given in \app{A-Solving-coeffs})
		\BE	\label{eq:Eq-Likelihoods-constraint}
			\hat{\ell} \left( \xx_i \right) = \ell \left( \xx_i \right) \,.
		\EE
	\item Finally, we compute the set of values of the model parameters optimizing the approximated 
		log-likelihood {function} using the DFPMIN IDL routine and take it as the maximum-likelihood estimator 
		of the set of values optimizing the exact log-likelihood {function}. To ensure the reliability of the maximum 
		likelihood estimator (MLE; see \sect{S-Likelihood-single-maxima}) obtained with DFPMIN, 
		we apply the latter from (i) the point of the design that possesses the largest likelihood 
		{function value} prior 
		to step (4), (ii) $N^d$ points spanning the entire parameter space and where $N$ 
		{($\ne n$)} is a relatively 
		low number of points (typically $N \lesssim 10$), and (iii) the likelihood-weighted average 
		position of these $N^d$ points (\ie their center of mass). Starting from these $N^d + 2$ points 
		ensures that at least one of them will lead DFPMIN to converge towards the global maximum 
		when the approximated log-likelihood {function} contains multiple global and local maxima.
\end{enumerate}

An RBF is a real-valued function that only depends on the Euclidean distance to a center whose location 
can be set arbitrarily. RBFs provide a class of functions that possess particularly interesting properties 
such as continuity, smoothness, and infinite differentiability. Their use is widely spread in various branches 
of applied mathematics and computer science including, \eg function approximation \citep{Powell93,Buhmann03}, 
data mining and interpolation \citep{Harder72,Lam83,Nychka15}, numerical analysis with meshfree 
methods for, \eg solving partial differential equations in numerical simulations 
\citep{Fasshauer07,Flyer11,Fornberg15,Flyer16}, computer graphics and machine learning 
\citep{Broomhead88,Boser92}. Polyharmonic Splines (PHS) are a type of infinitely smooth RBFs 
that does not possess any free parameter requiring a manual tuning. PHS can therefore be easily 
implemented for automated calculations.

As previously stated, the goal behind combining sparse calculations of a log-likehood 
with an RBF decomposition is to 
{limit} 
the number of model evaluations / forward 
calculations ($n$) to reduce the computational cost while maintaining a good accuracy 
on retrieving the exact maximum likelihood. 
{Through} 
low number of model evaluations, we mean to keep 
$n \lesssim 100-300$ {\it regardless of the dimension of the parameter space}, such that all model 
evaluations can easily be performed at once in parallel on a high-performance computing cluster. 
This provides us with a significant advantage as compared with more traditional 
{sequential optimization} 
methods 
since the effective computational time of our optimization method would only correspond 
to the computational time of {\it one} model evaluation (because steps 4 and 5 of the method 
only take up to $\lesssim 30$ seconds as long as $n \lesssim 500$). The optimization method 
we propose would, in general, also be more advantageous than a full grid search. Indeed, 
an accurate full grid search would typically require to sample each parameter of the $d$-space 
with about $50-100$ points at the least. This rapidly leads to a number of model evaluations 
that 
{is not practical even when using parallel computing.} 
Finally, ROAM should 
{be competitive} 
with genetic algorithms. Genetic algorithms 
applied to small population samples, \eg $\lesssim$ a few 100 points, typically require on the order 
of hundred generations to converge \citep[\eg][and references therein]{Louis92,Gibson98}, 
while faster convergence would require larger population sets. For ROAM, the equivalent 
of a population sample is a design of the parameter space and the equivalent of a generation 
would be an iteration of ROAM on a smaller parameter space region. For a population/design 
of $n$-points, ROAM should, in principle, be able to converge towards the solution without 
the need for iterations and, hence, 
{we estimate} 
would be at least 50-100 times faster than a genetic 
algorithm with the same population/design. In practice, preliminary tests of an iterative 
implementation of ROAM, which will be published in a subsequent paper, show robust 
and accurate convergence of ROAM within a few iterations (typically $< 10$).


\section{Results} \label{sec:S-Results}

In this section, we present a set of test cases performed on a synthetic test bed 
to validate ROAM (\sect{S-Method}) prior to any observational application. The set 
of test cases aims at assessing the performance of our method in different circumstances 
and defining a framework of application that will make use of its strengths.


\subsection{Numerical setup for the forward calculations} \label{sec:S-Setup}

\begin{figure}
   \centerline{\includegraphics[width=0.9\textwidth,clip=]{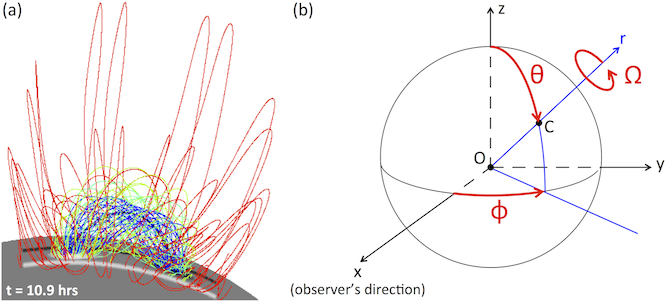}
              }
              \textbf{\refstepcounter{figure}\label{fig:Fig-FR-3paras} Figure \arabic{figure}.}{ \textbf{(a)} 3D view of the magnetic field of our synthetic test bed. The gray color scale display the photospheric magnetic flux (black/white for negative/positive magnetic flux). The green and blue lines show the magnetic field lines of the twisted flux rope. The red lines correspond to the magnetic field lines of the embedding magnetic field. \textbf{(b)} Schematic of the 3 parameters considered for our first study. The black thin solid lines highlight the solar photosphere. $(\theta, \phi)$ correspond to the angular coordinates of $C$, the photospheric center of the 3D box containing the magnetic field of our test bed, while $\Omega$ is the rotation angle of that 3D box around the solar radial direction passing by $C$.}
   \end{figure}

Our goal is to use the proposed optimization method for data-constrained modeling 
of the solar coronal magnetic field using, in particular, coronal polarimetric observations 
{(\ie the four Stokes parameters, $(I, Q, U, V)$, where Stokes $I$ is the total line 
intensity, Stokes $V$ is the circular polarization, and Stokes $Q$ and $U$ are the two 
components of the linear polarization).} 
All our test cases are therefore applied to a 3D model of magnetic fields chosen to represent 
scenarios typically observed in the solar atmosphere. The considered magnetic model is 
that of a 3D coronal magnetic flux rope generated from a 3D MHD numerical simulation 
of the emergence of twisted magnetic fields in the solar corona \citep[panel (a) of 
\fig{Fig-FR-3paras};][]{Fan12}.

\begin{figure}
   \centerline{\includegraphics[width=0.98\textwidth,clip=]{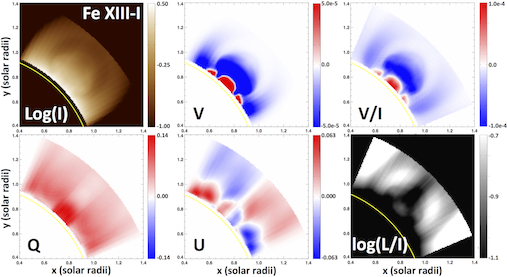}
              }
              \textbf{\refstepcounter{figure}\label{fig:Fig-3paras-GT-FS} Figure \arabic{figure}.}{ Coronal synthetic images of the {polarization signal} for the ground truth. {All four Stokes parameters ($I$, $Q$, $U$, $V$) are displayed together with the percentage of circular ($V/I$) and linear ($L/I=\sqrt{(Q^2 + U^2)}/I$) polarization.} The yellow solid line shows the solar limb.}
   \end{figure}

For the test cases, this magnetic field is assumed to depend on four parameters, \ie height 
in the corona 
{($h$; monotonically depends on the time of the MHD simulation, though not linearly),} 
co-latitude ($\theta$), longitude ($\phi$), and rotation angle\footnote{{Note that $(\theta; \phi; \Omega)$ 
are the co-latitude, longitude, and rotation angle of the numerical box -- containing the magnetic field 
of the MHD simulation -- around the Sun, while $h$ is the actual height of the flux rope in that numerical box 
(inside of which the solar photosphere is located at $h=0$).}} 
($\Omega$; panel (b) of \fig{Fig-FR-3paras}). 
A series of synthetic {polarimetric} data, referred to as the ground truth (GT) in the following, 
is generated for the flux rope associated with 
$(h; \theta; \phi; \Omega) = (0.16 \ \mathrm{R}_{\odot}; 45^{\circ}; 90^{\circ}; 30^{\circ})$ 
(see \fig{Fig-3paras-GT-FS}; note that both {Stokes} $Q$ and $U$ are presented in a frame of reference 
relative to the local vertical, or radial coordinate). All synthetic data are computed using 
the FORWARD Solar Soft  IDL package with a field-of-view (FOV) set to 
$y \times z = [0 \ \mathrm{R}_{\odot} ; 1.5 \ \mathrm{R}_{\odot}]^2$ (where $y$ and $z$ are 
the POS coordinates) and $x = [-0.79 ; 0.79] \ \mathrm{R}_{\odot}$ for the LOS. We use 
192 points along both directions for the POS and 80 points for each LOS, leading to spatial 
resolutions of $7.6''$ and $19.3''$ respectively. We limit the forward calculations of 
the polarization signals to a radial range of $[1.03 ; 1.5] \ \mathrm{R}_{\odot}$, \ie the FOV 
of CoMP. Although the spatial resolution of CoMP is $4.5''$, we restrict ourselves to a spatial 
resolution of $7.6''$ to allow for relatively fast (about 4-5 minutes on a MacBook Pro with a 2.7 GHz 
Intel Core i7 processor) calculations of the polarization signals while maintaining a quasi-CoMP 
resolution. We impose this FOV and POS spatial resolution to show that CoMP data currently 
carry meaningful information that can be used to constrain 3D reconstructions of the solar 
coronal magnetic field.

Finally, it should be emphasized that the considered flux rope possesses a strong degree 
of symmetry, such that 
$\bb (\theta; \phi=90; \Omega \pm 180^{\circ}) = - \bb (\theta; \phi=90; \Omega)$. 
We will exploit these symmetry properties to test ROAM when faced with a log-likelihood {function} 
containing multiple maxima.

\subsection{Likelihood function with a single maximum} \label{sec:S-Likelihood-single-maxima}

We first apply ROAM in the context of a 3D likelihood {function} possessing a single 
maximum. The parameters considered for this study are the co-latitude, longitude, and rotation 
angle, \ie $(\theta; \phi; \Omega)$. We then build a likelihood {function} that takes into account all four 
Stokes parameters, \ie $I$, $Q$, $U$, and $V$. For the set $\{ \xx_i \}$ of a design, we first 
define the log-likelihood {function} for a given Stokes parameter, $S=\{I,Q,U,V\}$, up to a constant, as
	\BE	\label{eq:Eq-Likelihood-Stokes}
		\ell_{S} (\xx_i) = f \left( S (\xx_i) - S_{gt} \right) = - \sum_{k} \left( S_{k} (\xx_i) - S^{gt}_{k} \right)^2 \,,
	\EE
where $k$ is the $k$-th pixel of the Stokes, $S$, image. The final log-likelihood {function} is then constructed as
	\BE	\label{eq:Eq-Likelihood-3paras}
		\ell (\xx_i) = w_I \ell_I (\xx_i) + w_Q \ell_Q (\xx_i) + w_U \ell_U (\xx_i) + w_V \ell_V (\xx_i)  \,,
	\EE
where the weighting coefficients $w_S$ were chosen to ensure that $I$, $Q$, $U$, and $V$ 
similarly contribute to the log-likelihood {function}, which behavior would otherwise be dominated by 
the quantity possessing the largest values (here, Stokes $I$). We use 
$\left( w_I ; w_Q ; w_U ; w_V \right) = \left( 1.3 \times 10^{-4} ; 1.9 \times 10^{-2} ; 9.2 \times 10^{-2} ; 1.2 \times 10^4 \right)$.

With the log-likelihood {function} defined in \eq{Eq-Likelihood-3paras}, we consider 3 test cases referred 
to as 3DN31, 3DN301, and 3DN31ZOOM (see \tab{Tab-3paras-test-beds}). These 3 test cases 
each contain 100 different designs and differ by the number of points in the designs (31 or 301) 
as well as by the size of the parameter space to allow us to investigate their role on the performances 
of ROAM. These test cases are designed to allow us to determine the criteria required 
for the method to ensure robustness and reliability of the results, \ie such that {the method} provides 
a maximum likelihood estimator (MLE) that gives a good approximation of the parameters of 
the maximum of the exact likelihood {function} independently of the design and number of points used.

\begin{table}[!t]
\textbf{\refstepcounter{table}\label{tab:Tab-3paras-test-beds} Table \arabic{table}.}{ Characteristics of the test with a likelihood {function} possessing a single maximum. }

\processtable{ }
{\begin{tabular}{llllllll}\toprule
 & $n$ & $t_{\mathrm{elapsed}}$ (hrs) & $t_{\mathrm{full}}$ (hrs) & h ($\mathrm{R}_{\odot}$) & $\theta$  ($^{\circ}$) & $\phi$  ($^{\circ}$) & $\Omega$  ($^{\circ}$) \\\midrule
3DN31 & 31 & 2.6 & $2.5 \times 10^3$ & 0.16 & $[24; 66]$ & $[60; 120]$ & $[0; 90]$ \\
3DN301 & 301 & 25 & $2.3 \times 10^6$ & 0.16 & $[24; 66]$ & $[60; 120]$ & $[0; 90]$ \\
3DN31ZOOM & 31 & 2.6 & $2.5 \times 10^3$ & 0.16 & $[42; 48]$ & $[75; 105]$ & $[15; 45]$ \\\botrule
\end{tabular}}{{\textbf{Note:}} $n$ is the number of points per design. $t_{\mathrm{elapsed}}$ is the elapsed time for forwarding the Stokes images associated with the $n$ points of a design in series, while $t_{\mathrm{full}}$ is the total elapsed time that would be required to compute Stokes images for the $n^3$ points of the 3D parameter space in series. Each test case contains 100 randomly-chosen different designs. The naming convention is such that ``xD'' indicates the dimension of the parameter space and ``Nx'' indicates the number of points per design ($n$). {The polarimetric data for the ground-truth are associated with $(h; \theta; \phi; \Omega) = (0.16 \ \mathrm{R}_{\odot}; 45^{\circ}; 90^{\circ}; 30^{\circ})$.}}
\end{table}

For each test case, the parameters of the RBF decomposition are 
{$d = 3$, $m = 3$ and $p = m - 1 = 2$.} 
We choose the minimum $m$ satisfying the condition $2m - d > 1$ (see \sect{S-Method}). 
Although $\theta$, $\phi$, and $\Omega$ all are periodic parameters, their corresponding 
range is smaller than half the associated period and, hence, no periodic effect is expected. 
As explained \app{A-Periodic-Variables}, disregarding the periodicity and curvature of 
the $d$-space should not significantly affect the results in such circumstances. We therefore 
ignore the periodicity of $\theta$, $\phi$, and $\Omega$ in all 3D cases considered in this section, 
but return to the issue of periodicity in \sect{S-Likelihood-multiple-maxima}.

\begin{figure}
   \centerline{\includegraphics[width=0.98\textwidth,clip=]{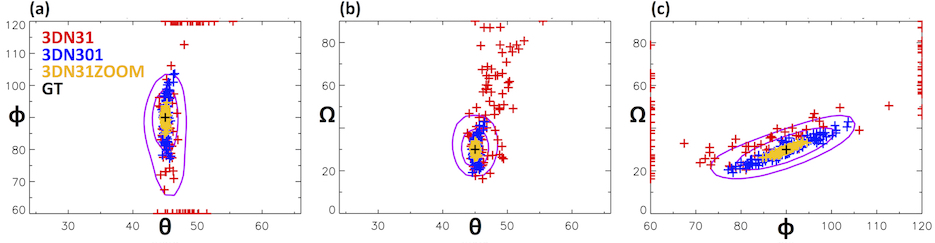}
              }
              \textbf{\refstepcounter{figure}\label{fig:Fig-3paras-dispersions} Figure \arabic{figure}.}{ 2D scatter plots of the maximum likelihood estimators (MLEs) found for each design of the 3DN31 (red crosses), 3DN301 (blue crosses), and 3DN31ZOOM (yellow crosses) cases. The black cross highlights the position of the exact maximum (\ie ground-truth). The two purple solid lines show $[0.9;0.95] \times \max (\ell)$ isocontours. Note that, in panel (b), the red crosses (MLEs of 3DN31) outside the two log-likelihood {function} isocontours are the solutions associated with $\phi_{\mathrm{MLEs}} = 60^{\circ}$ and $\phi_{\mathrm{MLEs}} = 120^{\circ}$ from panels (a) and (c).}
   \end{figure}

\fig{Fig-3paras-dispersions} presents 2D dispersion plots of the MLEs obtained for each one 
of the 100 randomly-chosen designs of the 3DN31 (red), 3DN301 (blue), and 3DN31ZOOM 
(yellow) cases. For the 3DN31, the MLEs are fairly weakly dispersed for the $\theta$ parameter, 
spanning a range of roughly $10^{\circ}$. As summarized in \tab{Tab-3paras-results}, 
the root mean square (hereafter, rms) of the MLEs, $\theta_{\mathrm{rms}}$, is $\approx 47.1^{\circ}$, 
which is only $\approx 2.1^{\circ}$ different from $\theta_{\mathrm{GT}} = 45^{\circ}$. This suggests 
that $\theta_{\mathrm{MLEs}}$ is not overly sensitive to the design used for the RBF decomposition. 
These conclusions contrast with both the $\phi$ and $\Omega$ parameters. 
Although $\phi_{\mathrm{rms}} \approx 92.8^{\circ}$ is very close to the ground-truth, 
$\phi_{\mathrm{GT}}=90^{\circ}$, the ensemble of solutions, $\phi_{\mathrm{MLEs}}$, spans 
the entire $\phi$-range considered for the 3DN31. Similarly poor results are obtained for the set 
of $\Omega_{\mathrm{MLEs}}$, whose rms is $\approx 21^{\circ}$ off from the ground-truth, 
{$\Omega_{\mathrm{GT}} = 30^{\circ}$} (see \tab{Tab-3paras-results}). 
{\fig{Fig-3paras-dispersions} further shows that there is a strong coupling between $\phi$ 
and $\Omega$. In particular, we find that $\Omega_{\mathrm{MLEs}}$ provide a poor estimation 
of $\Omega_{\mathrm{GT}}$ whenever $\phi_{\mathrm{MLEs}}$ are themselves a poor estimation 
of $\phi_{\mathrm{GT}}$ (and vice-versa).} 
Such results trace very poor performances of our optimization 
method for the chosen setup of the 3DN31 case. The MLE strongly depends on the design used 
to perform the RBF decomposition. Hence, the MLE obtained from applying our method to a single 
design is not reliable for the setup of the 3DN31 case.

When comparing 3DN31 to 3DN301 in \fig{Fig-3paras-dispersions}, we can see that increasing 
the number of points significantly improves the performances of ROAM (see dark blue crosses). 
For all three parameters, the rms is only $\approx 0.2^{\circ}$ off the ground-truth for 3DN301. 
The range spanned by the ensemble of solutions is 
{relatively} 
smaller than for 3DN31, 
{\it $\approx 5$ times smaller} for $\theta_{\mathrm{MLEs}}$ and {\it $\approx 3$ times smaller} 
for both $\phi_{\mathrm{MLEs}}$ and $\Omega_{\mathrm{MLEs}}$. 
{Again, the strong coupling between $\phi$ and $\Omega$ is still present but its effect 
on their uncertainties is strongly reduced as compared with the case 3DN31.} 
Such results are 
not yet perfect since the set of solutions for both $\phi$ and $\Omega$ is spread over $20^{\circ}$, 
which is relatively significant considering the range of their values. However, they do show that 
increasing the number of points per design strongly helps in reducing (a) the dependence of the MLE 
on the design used for the RBF decomposition, 
{ and (b) the effect of (even strong) coupling of parameters on their uncertainty. 
Increasing the number of points per design therefore strongly helps } 
in improving the reliability and robustness 
of the proposed optimization method.

\begin{table}[!t]
\textbf{\refstepcounter{table}\label{tab:Tab-3paras-results} Table \arabic{table}.}{ Optimization results for a likelihood {function} with a single maximum. }

\processtable{ }
{\begin{tabular}{lllllll}\toprule
 & & $ \theta_{\mathrm{rms}}$ & & $\phi_{\mathrm{rms}} $ & & $ \Omega_{\mathrm{rms}} $ \\\midrule
3DN31 & & $47.1$ & & $92.8$ & & $51.0$ \\
3DN301 & & $45.2 $ & & $90.0 $ & & $30.2$ \\
3DN31ZOOM & & $45.0$ & & $89.8$ & & $29.7$ \\\botrule
\end{tabular}}{{\textbf{Note:}}  {The ground-truth parameters are $(\theta; \phi; \Omega) = (45^{\circ}; 90^{\circ}; 30^{\circ})$.} All angles are in degrees.}
\end{table}

Compared with 3DN31, the 3DN31ZOOM case is used to investigate the effect of focusing 
the parameter space around a region closer to the exact maximum while keeping the number 
of points per design constant. \fig{Fig-3paras-dispersions} shows that reducing the size 
of the parameter space is also beneficial for reducing the MLEs dispersion (see yellow crosses). 
The rms values for all three parameters are as close to the ground-truth as for 3DN301 
(see \tab{Tab-3paras-results}) and the solutions are spanning a range that is 
{\it $\approx 4$ times smaller} than for 3DN301 and {\it $\approx 12$ times smaller} than 
for 3DN31. The MLEs are now independent of the design used for the RBF decomposition 
for $\theta$ and very weakly dependent on that design for both $\phi$ and $\Omega$. 
{The effect of the strong $\phi-\Omega$ coupling on their uncertainty is again strongly 
reduced and even smaller than for the 3DN301 case.} 
These very good results prove that ROAM can perform very well and provide an accurate 
estimation of the ground-truth parameters when the setup is suitably defined.

  \begin{figure}
   \centerline{\includegraphics[width=0.8\textwidth,clip=]{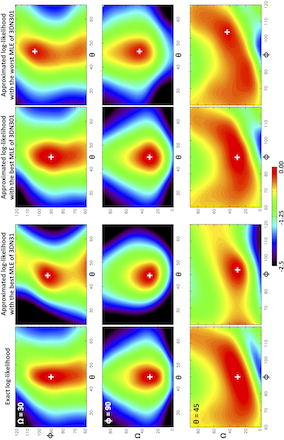}
              }
              \textbf{\refstepcounter{figure}\label{fig:Fig-3paras-all3D-LfS-1} Figure \arabic{figure}.}{ Comparison between approximated and exact log-likelihood {functions}. The white ``+'' symbol indicates the position of the maximum log-likelihood.}
   \end{figure} 

\fig{Fig-3paras-all3D-LfS-1} displays 2D cuts of the exact log-likelihood {function} and the approximated 
ones associated with the designs of 3DN31 and 3DN301 giving the best MLEs (referred 
to as {\it best cases} in the following), as well as the approximated log-likelihood {function} associated 
with the design of 3DN301 giving the worst MLEs (referred to as {\it worst case}). Here, 
the best (worst) MLE is defined as the MLE minimizing (maximizing) the distance to the ground 
truth in the parameter space. The best MLE from 3DN31 is 
$( \theta ; \phi ; \Omega)_{\mathrm{best-MLE}} = ( 44.6^{\circ} ; 92.9^{\circ} ; 30.1^{\circ} )$ 
while the best MLE from 3DN301 is 
$( \theta ; \phi ; \Omega)_{\mathrm{best-MLE}} = ( 45.0^{\circ} ; 89.9^{\circ} ; 30.3^{\circ} )$. 
The worst MLE from 3DN301 is 
$( \theta ; \phi ; \Omega)_{\mathrm{best-MLE}} = ( 46.5^{\circ} ; 103.7^{\circ} ; 43.0^{\circ} )$. 
The figure shows that the approximated log-likelihood {function} of the best case from 3DN31 gives 
an overall rough approximation of the exact one both in terms of values and shape. 
Note, though, that the rms error on the log-likelihood is $0.24$, which is rather small given 
that $\max ( | \ell (\xx) | ) \approx 3.5$ for the considered parameter space. For the log-likelihood {function} 
of the best case from 3DN301, the results are very much better. The approximated log-likelihood {function} 
is able to accurately capture both the values and shape of the exact log-likelihood {function}; the rms error is 
$0.05$, \ie $\approx 5$ times smaller than for the best case of 3DN31. For the worst case of 
3DN301, the rms error on the log-likelihood is $0.25$, which is very similar to that of the best 
case of 3DN31, and the MLE is 
{far from} 
the ground truth for both $\phi$ and $\Omega$. 
However, we find that the worst case from 3DN301 provides a more accurate RBF decomposition 
of the exact log-likelihood {function} than the best case of 3DN31; the log-likelihood {function} surfaces display 
a similar pattern as for the best case of 3DN301 but shifted in the $\Omega$ direction. 
The difference with the 3DN31 lies in the density of points in the entire design, and in the vicinity 
of the exact maximum, with regard to the structuring, or gradients, of the exact log-likelihood {function}. 
This is because the goodness of the approximation is determined by that of the RBF decomposition, 
which depends on the number of constraints - and hence, points - brought by the design. 
In other words, the more structured the exact log-likelihood {function}, the stronger the effect of point 
density on the goodness of the RBF decomposition / log-likelihood {function} approximation, similar 
to what one would expect when discretizing a continuous functions that contains strong 
gradients. While not shown here, the combined effect of point density and log-likelihood {function} 
structuring on the quality of the RBF decomposition is further supported and illustrated 
by the best case of 3DN31ZOOM that provides the best approximation of the exact log-likelihood {function} 
in the vicinity of the exact maximum even though the corresponding design only includes 31 points.

The aforementioned results show that the RBF decomposition performed 
{well} 
from a sparse sampling of the parameter space, and hence ROAM, is able to capture 
both the values and variations of the exact log-likelihood {function} when suitable conditions are met, 
namely, when the design contains a high enough density of points in the surroundings of 
the exact maximum and in areas where the exact log-likelihood {function} is strongly structured. 
They further demonstrate that ROAM can perform well even with a very low number 
of points per design although not as robustly. The combined results from 3DN31 and 
3DN31ZOOM indicate that an iterative application of ROAM with a smaller and smaller 
parameter space would be an interesting way to improve its robustness when used 
with a very sparse design. Such a robust approach has been successfully tested 
{but is beyond the scope of this work} 
and will be presented in a subsequent paper. {In particular, the iterative implementation 
of ROAM strongly improves $\phi_{\mathrm{MLEs}}$ and $\Omega_{\mathrm{MLEs}}$, 
leading to a better than $0.5^{\circ}$ accuracy on both of these parameters in typically 
4-5 iterations with designs of 31 points. The solution is quasi-independent of the design 
used for the RBF decomposition and the strong $\phi-\Omega$ coupling (previously 
mentioned and visible in panel c of \fig{Fig-3paras-dispersions} and in the $\phi-\Omega$ 
cut of the exact log-likelihood function shown in \fig{Fig-3paras-all3D-LfS-1}) is comfortably 
reduced and overcome (note that such coupling could also be overcome by separately 
optimizing one of the coupled parameters, \eg $\Omega$, and apply ROAM to the 2D 
parameter space $(\theta ; \phi)$).}

  \begin{figure}
   \centerline{\includegraphics[width=0.98\textwidth,clip=]{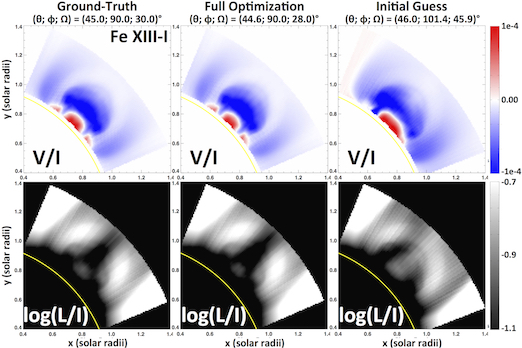}
              }
              \textbf{\refstepcounter{figure}\label{fig:Fig-3paras-n301-VLOI-RFoIg} Figure \arabic{figure}.}{ Comparison between polarization signal showing the benefit of steps 4 and 5 of ROAM (see \sect{S-Method}), for one of the designs of the 3DN301 case. Left column: ground-truth. Middle Column: fully optimized solution (all five steps of the method are applied). Right column: initial guess from the design (i.e., when omitting steps 4 and 5 of ROAM).}
   \end{figure} 

Note that, with the goal of increasing speed, and considering our previous comments on 
the acceptable degree of roughness in the log-likelihood {function} approximation and an iterative 
implementation of ROAM, a less sophisticated approach might be conceived. For instance, 
one could first go through steps 1 to 3 of the method (see \sect{S-Method}). Then, step 4 
(\ie the RBF decomposition) would be replaced by taking the point of the design associated 
with the highest log-likelihood {function value} as a temporary MLE and one would iterate the procedure by 
defining a smaller design centered around the temporary MLE until a convergence criterion 
is reached. There are several reasons for not making such a choice. The main reason is that 
such an initial guess can be far from the exact maximum likelihood, which would likely slow down 
the convergence by requiring unnecessary iterations and would make the final result more 
sensitive to local maxima. In addition, applying the RBF decomposition and the search for 
the maximum from the approximated log-likelihood {function} is computationally cheap when the number 
of RBFs is as small as for the cases considered in this study, \ie typically takes less than 10 
seconds for the designs of the 3DN301 case. The benefits of applying steps 4 and 5 of 
ROAM as proposed in \sect{S-Method} (\ie the RBF decomposition and 
the search for the MLE from the RBFs approximated log-likelihood {function}) are illustrated in 
\fig{Fig-3paras-n301-VLOI-RFoIg}. The figure displays Stokes images for the ground truth, 
the MLE obtained from fully applying our optimization method to one design of the 3DN301, 
and for the initial guess from that design. As one can see, the initial MLE guess from the design, 
$( \theta ; \phi ; \Omega)_{\mathrm{IG}} = ( 46.0^{\circ} ; 101.4^{\circ} ; 45.9^{\circ} )$, has a $\phi_{\mathrm{IG}}$ 
and $\Omega_{\mathrm{IG}}$ that are far off both the ground truth 
($( \theta ; \phi ; \Omega)_{\mathrm{GT}} = ( 45.0^{\circ} ; 90.0^{\circ} ; 30.0^{\circ} )$) 
and the MLE obtained from the RBF decomposition (full application of ROAM; 
$( \theta ; \phi ; \Omega)_{\mathrm{MLEs}} = ( 44.6^{\circ} ; 90.0^{\circ} ; 28.0^{\circ} )$). These strong 
differences in $\phi$ and $\Omega$ result in significantly different Stokes profiles. Iterations 
would then be needed for the results to be as close to the ground truth as the MLE from 
the full optimization, which (1) gives a very good estimation of the parameters of the exact 
maximum likelihood without any real need for iterations, and (2) only takes a few more 
seconds of calculations.

  \begin{figure}
   \centerline{\includegraphics[width=0.98\textwidth,clip=]{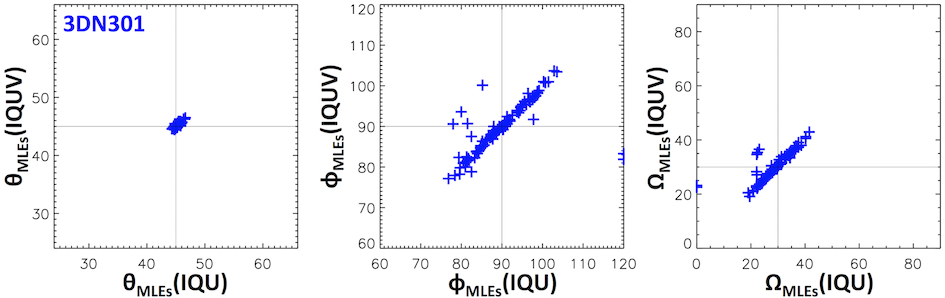}
              }
              \textbf{\refstepcounter{figure}\label{fig:Fig-3paras-n301-dispersions-IQUV-V-vs-IQU} Figure \arabic{figure}.}{ Scatter plots showing the effect of using the circular polarization signal on the MLEs for the 3DN301 case. The thin black solid lines indicate the value of the ground truth.}
   \end{figure} 

In practice, the current capabilities of the CoMP instrument and calibration software do 
not allow routine measurements of Stokes $V$ since the signal-to-noise ratio is too small. 
We therefore perform an additional test to show that the current linear polarization signal 
from CoMP is sufficient to constrain the parameters of a magnetic model using ROAM. 
The log-likelihood {function} is defined as in \eq{Eq-Likelihood-3paras} keeping $w_I$, $w_Q$, 
$w_U$ as before, but now setting $w_V = 0$. The results of that study are displayed in 
\fig{Fig-3paras-n301-dispersions-IQUV-V-vs-IQU} for 3DN301. The figure presents scatter plots 
of the MLEs of a design obtained when using all four Stokes versus obtained when using Stokes 
$I$, $Q$, and $U$ only. In such plots, the points should form a line of equation $y=x$ whenever 
the solutions obtained one way or the other remain the same. As one can see from 
\fig{Fig-3paras-n301-dispersions-IQUV-V-vs-IQU}, this is exactly the case for $\theta_{\mathrm{MLEs}}$. 
Most of the points are also forming a straight line, $y=x$, for both $\phi_{\mathrm{MLEs}}$ and $\Omega_{\mathrm{MLEs}}$, 
with only about 7-8 points (out of 100) being off the line. Such results indicate that a log-likelihood {function} 
built from {Stokes} $I$, $Q$, and $U$ contains sufficient information to constrain the three spatial location 
and orientation parameters considered here. We therefore conclude that the current linear polarization 
measurements from CoMP contain sufficient observational information to constrain {\it some} 
of the parameters of a given magnetic model.


\subsection{likelihood function with multiple maxima} \label{sec:S-Likelihood-multiple-maxima}

\begin{table}[!t]
\textbf{\refstepcounter{table}\label{tab:Tab-2paras-test-beds} Table \arabic{table}.}{ Characteristics of the test with a likelihood {function} possessing multiple maxima. }

\processtable{ }
{\begin{tabular}{llllllll}\toprule
 & $n$ & $t_{\mathrm{elapsed}}$ (hrs) & $t_{\mathrm{full}}$ (hrs) & h ($\mathrm{R}_{\odot}$) & $\theta$ ($^{\circ}$) & $\phi$ ($^{\circ}$) & $\Omega$ ($^{\circ}$) \\\midrule
2DN120 & 120 & 10 & $1.2 \times 10^3$ & $[0.04;0.52]$ & $45$ & $90$ & $[0; 357]$ \\\botrule
\end{tabular}}{{\textbf{Note:}} The test case contains 100 different designs. $t_{\mathrm{full}}$ is the total elapsed time that would be required to compute the Stokes images for the $n^2$ points of the 2D parameter space in series. {The polarimetric data for the ground-truth are associated with $(h; \theta; \phi; \Omega) = (0.16 \ \mathrm{R}_{\odot}; 45^{\circ}; 90^{\circ}; 30^{\circ})$.}}
\end{table}

In this section, we test ROAM in the case of a log-likelihood {function} with multiple maxima having 
similar values. For that purpose, we only build the log-likelihood {function} with Stokes $Q$ and $U$, 
setting the weight coefficients of \eq{Eq-Likelihood-3paras} to 
$(w_I; w_Q, w_U, w_V) = (0.; 2.0 \times 10^{-2} ; 6.9 \times 10^{-2} ; 0.)$. Only the height 
of the flux rope in the corona, $h$, and the tilt angle, $\Omega$, are considered for this test 
(see \tab{Tab-2paras-test-beds} for the range of values considered for each parameter).

\begin{figure}
   \centerline{\includegraphics[width=0.9\textwidth,clip=]{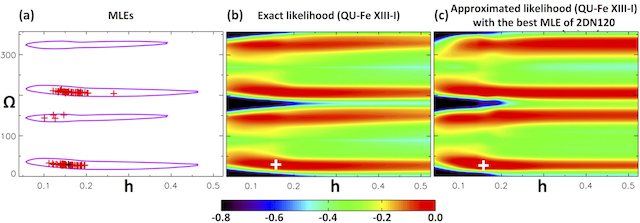}
              }
              \textbf{\refstepcounter{figure}\label{fig:Fig-2paras-LfS-RB-dispersions} Figure \arabic{figure}.}{ likelihood {function} with multiple maxima. (a) Scatter plots of the maximum likelihood estimators (MLEs) found for each design of the 2DN120 (red crosses) case. (b) {Surface plot of an} exact log-likelihood {function} possessing 2 global ($\Omega = \{30; 210\}^{\circ}$) and 2 local ($\Omega = \{150; 330\}^{\circ}$) maxima. (c) Approximated log-likelihood {function} with the best MLE of 2DN120. The white and black crosses highlight the position of the exact maximum (\ie ground-truth). The purple solid lines show $0.95 \times \max (\ell)$ isocontours.}
   \end{figure}

Stokes $Q$ and $U$ signals are associated with the transverse magnetic field, \ie the component 
of a magnetic field perpendicular to the LOS. For a single point in the solar corona, the transverse 
magnetic field diagnosed from either the Hanle or Zeeman effect is subject to a $180^{\circ}$ 
ambiguity \citep[\eg][]{Casini99,Judge07}. In terms of the parameters considered 
in our tests, it means that a single point magnetic field set with a rotation angle, $\Omega_{\mathrm{SP}}$, 
will give the same Stokes $Q$ and $U$ signals as when set with $\Omega_{\mathrm{SP}} \pm 180^{\circ}$. 
Considering that $\phi_{\mathrm{GT}} = 90^{\circ}$ (that is, the flux rope is centered at the solar limb) 
and the strong symmetry of our flux rope (see \sect{S-Setup}), we expect the LOS integrated 
Stokes $Q$ and $U$ to be the same for $\Omega$ and $\Omega \pm 180^{\circ}$, resulting 
in a log-likelihood {function} with two maxima respectively located at $\Omega_{\mathrm{GT}}$ and 
$\Omega_{\mathrm{GT}} \pm 180^{\circ}$; note that the symmetry of Stokes $Q$ and $U$ would be broken 
if the flux rope were not centered on the solar limb. This is indeed the case as shown 
in panel (b) of \fig{Fig-2paras-LfS-RB-dispersions} where a maximum region can be 
observed at $\Omega = 30^{\circ}$ and $\Omega = 210^{\circ}$. Panel (b) of 
\fig{Fig-2paras-LfS-RB-dispersions} further shows the presence of two additional maximum 
regions located at $\Omega = 150$ and $330^{\circ}$. These two solutions suggest a symmetry 
with regard to the plane $\Omega = 0^{\circ}$ that is not expected. We find that the corresponding 
{Stokes} $Q$ and $U$ images are, as expected, different from those of the ground truth. However, the differences 
are small as compared with other values of $\Omega$, resulting in a local maximum in those 
two regions. Note, though, that these four maximum regions are only possible because 
$\phi = 90^{\circ}$, whereas any other value of $\phi$ would break the symmetry of the Stokes $Q$ 
and $U$ images.

In the present test case, the periodic parameter $\Omega$ varies on a range of values larger 
than half its period. In such circumstances, we must consider its periodicity for the RBF 
decomposition (see \app{A-Periodic-Variables}). Accordingly, the parameters of the RBF 
decomposition are 
{$d' = 3$, $m = 3$ and $p = m - 1 = 2$}. 
As in \sect{S-Likelihood-single-maxima}, we run 
our optimization method on 100 different designs whose properties are given in 
\tab{Tab-2paras-test-beds}. The results are summarized in a 2D dispersion plot in panel (a) 
of \fig{Fig-2paras-LfS-RB-dispersions}. The figure shows that the 100 MLEs are mainly, 
and almost equally, clustering around the two global $\Omega$ maximum regions, corresponding 
to the ground truth and its counterpart at $180^{\circ}$. We further find that, out of these 100 
solutions, only 4 are associated with one of the two local maximum regions, here 
$\Omega \approx 150^{\circ}$. As for the height of the MLEs, we find an average value of 
$1.6 \times 10^{-1} \ \mathrm{R}_{\odot}$ with a $2 \sigma$ dispersion level of 
$0.5 \times 10^{-1} \ \mathrm{R}_{\odot}$, meaning that the height is well constrained 
even from using Stokes $Q$ and $U$ only. The dispersion plot from panel (a) of 
\fig{Fig-2paras-LfS-RB-dispersions} therefore indicates that our optimization method is 
strongly sensitive to multiple global maxima and {\it can} be sensitive to local maxima. 
Note that the sensitivity to local maxima depends upon both the number of points used 
in the design and the value of these local maxima relatively to that of the global maxima.

Panels (c) of \fig{Fig-2paras-LfS-RB-dispersions} displays {a surface plot of} 
the log-likelihood {function} from 
the best case of 2DN120. As one can see, the RBF decomposition is able to capture both 
the values and shapes of the exact log-likelihood {function}. We find an rms error of $0.04$ 
on the log-likelihood. 
The RBF decomposition can therefore provide a good approximation of the exact log-likelihood {function} 
even with a periodic space and the presence of multiple maxima.

  \begin{figure}
   \centerline{\includegraphics[width=0.9\textwidth,clip=]{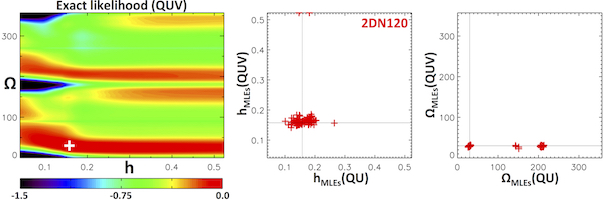}
              }
              \textbf{\refstepcounter{figure}\label{fig:Fig-2paras-LfS-QUV-vs-QU} Figure \arabic{figure}.}{ Effect of using circular polarization on a log-likelihood {function} with multiple maxima.}
   \end{figure} 

Finally, we show in \fig{Fig-2paras-LfS-QUV-vs-QU} that using the Stokes $V$ signal 
to build the log-likelihood {function} removes the $\Omega$ ambiguities that were observed 
in the log-likelihood {function} constructed from {Stokes} $Q$ and $U$ only. 
When using Stokes $V$, the optimization 
leads to $\Omega_{\mathrm{rms}} \approx 28.8^{\circ}$. This means that some additional 
observables might be worth considering to remove ambiguities in parameters when they exist. 
Another alternative to remove ambiguities is to reduce the parameter space to regions having 
a single maximum. Then, one can either study each region separately or use prior constraints 
to eliminate regions that are very unlikely. For instance, one can use the photospheric 
magnetograms, or H$\alpha$ observations, prior to or after the passage of the flux rope 
at a limb to estimate the rotation angle (\ie $\Omega$) and put strong constraints on the values 
of rotation angle to consider for the parameter space.


\subsection{Stability with regard to noise in the data} \label{sec:S-Stability-wrt-noise}

In practice, any real data is subject to measurement errors. Such errors may prevent 
the retrieval of any meaningful information about the polarization, and hence, the magnetic field 
in regions of weak signals and/or when the signal-to-noise ratio is weak. The results of 
ROAM might be sensitive to such noise and we therefore need to investigate that sensitivity. 
For that reason, we now test our method when the synthetic observations associated 
with the ground truth contain some noise. In this regard, we build the log-likelihood {function} 
with Stokes $U$ images only
	\BE	\label{eq:Eq-Likelihood-3paras-noise}
		\ell (\xx_i) = \frac{ \ell_U (\xx_i) }{ \sigma^{2}_{U}}   \,,
	\EE 
where $\sigma_U$ is the root mean square of the noise in the synthetic Stokes $U$ signal 
of the ground truth.

For a given value of photon noise, $\sigma_I$, $\sigma_Q$, $\sigma_U$, $\sigma_V$ are 
all different. As a consequence, if one uses more than one Stokes 
{component,} 
then varying the noise 
further changes the relative contribution of each Stokes parameter to the log-likelihood {function} 
due to the weighting by $1/\sigma_S$. We need to be free of the variation of relative 
contribution of the different Stokes in order to isolate the sole effect of noise 
on the robustness of our optimization method, which then implies using only one Stokes 
parameter to define the log-likelihood {function}. Considering CoMP capabilities and the current magnetic 
model and ground truth, we performed several tests with different levels of noise (which can be 
added using FORWARD) and found that (1) Stokes $V$ cannot be used for realistic exposure 
times because its values for our test bed are too weak and would require an unrealistic 4 days 
exposure time to reach a moderate level of noise for the particular choices of ground-truth 
parameters and pixel sizes (see corresponding values in \sect{S-Setup}), (2) Stokes $I$ cannot be 
used because 
{it is} 
not sensitive enough to noise (even a 1 second exposure time leads to a very weak 
level of noise), and (3) Stokes $Q$ and $U$ are better suited for the noise test with exposure times 
of the order of 1 to 100 seconds. From this analysis, we chose Stokes $U$ because it was slightly 
more sensitive to noise than Stokes $Q$ for the setup considered in this paper (note that both $Q$ 
and $U$ are presented in a frame of reference relative to the local vertical, or radial coordinate).

  \begin{figure}
   \centerline{\includegraphics[width=0.98\textwidth,clip=]{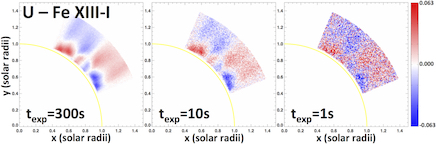}
              }
              \textbf{\refstepcounter{figure}\label{fig:Fig-3paras-GT-StokesU-noise} Figure \arabic{figure}.}{ Synthetic Stokes $U$ images of the ground truth for different exposure times, $t_{\mathrm{exp}}$, and hence, noise levels.}
   \end{figure} 

FORWARD already implements a photon noise calculation for the infra-red lines under 
consideration \citep[see \eg][]{Gibson16}. The noise is calculated according to the specifications 
of the instrument considered (telescope aperture, detector efficiency), the background photon 
level, and the exposure time to obtain a forward calculation that includes the noise. For CoMP, 
the aperture is 20 cm, the efficiency is 0.05 throughput and the background is 5 parts per million 
of solar brightness. We perform three tests with different exposure time, $t_{\mathrm{exp}}$, 
hence noise level, \ie $t_{\mathrm{exp}} = \left( 1 ; 10 ; 300 \right)$ seconds that respectively 
correspond to strong, moderate, and weak noise cases for the considered setup. The synthetic 
Stokes $U$ images of the ground truth for these noise levels are displayed in \fig{Fig-3paras-GT-StokesU-noise}. 
These synthetic ground truth are used with all designs of the 3DN31, 3DN301, and 3DN31ZOOM cases.

  \begin{figure}
   \centerline{\includegraphics[width=0.98\textwidth,clip=]{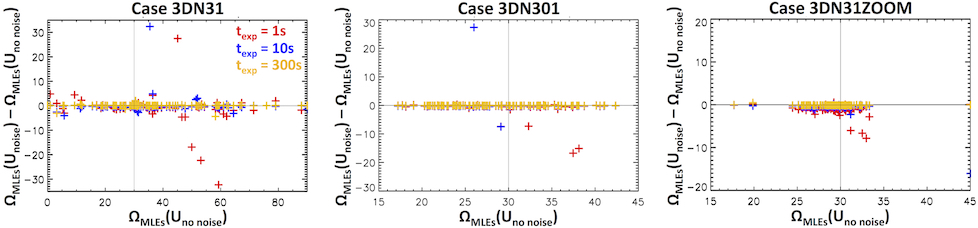}
              }
              \textbf{\refstepcounter{figure}\label{fig:Fig-3paras-dispersions-noise} Figure \arabic{figure}.}{ Scatter plots showing the effect of noise on the MLEs obtained with ROAM for all designs of 3DN31, 3DN301, and 3DN31ZOOM. The horizontal, thin black line indicates the zero error level, while the vertical thin black line indicates the ground-truth value.}
   \end{figure} 

\fig{Fig-3paras-dispersions-noise} presents scatter plots of the error on the MLEs 
obtained when noise is included in the ground truth Stokes $U$ images as compared 
with the case when no noise is considered. The plots are only shown for the $\Omega$ 
parameter because all three parameters $\theta$, $\phi$, and $\Omega$ display very 
similar results. In \fig{Fig-3paras-dispersions-noise}, one can see a nearly perfect 
horizontal line at $y=0$ for $\Omega_{\mathrm{MLEs}}$ obtained with an exposure time 
of 300 seconds (yellow crosses) for all test cases (3DN31, 3DN301, and 3DN31ZOOM). 
This means that the $t_{\mathrm{exp}} = 300$ seconds case is equivalent to the no-noise 
case. For cases $t_{\mathrm{exp}} = 10$ and $t_{\mathrm{exp}} = 1$ second, the plots show 
some departure from the $y=0$ line, which increases with the level of noise. The figure 
also shows that the noise effect on the robustness of the MLEs depends on the density 
of points in the designs, \ie 3DN31 is the most affected by the noise while 3DN31ZOOM is 
the least affected. That being said, we find that only less than $\approx 10-15$ points 
(out of 100) of 3DN31 exhibit a strong sensitivity to noise for the $t_{\mathrm{exp}} = 1$ 
second case, \ie with an error larger than $5^{\circ}$. This number drops to $\approx 5$ 
when $t_{\mathrm{exp}} = 10$ seconds. For the $t_{\mathrm{exp}} = 1$ second, 
\fig{Fig-3paras-GT-StokesU-noise} shows that the noise strongly masks the real Stokes $U$ 
signal, although not entirely. We therefore deduce that our optimization method is very stable 
against the presence of noise in the data as long as the noise does not entirely mask the real 
signal. Considering that Stokes $Q$ is similarly sensitive to noise as Stokes $U$ and that Stokes $I$ 
is much less sensitive to the noise, we conclude that our method can be robustly used with 
the Stokes $I$, $Q$, and $U$ data provided by the CoMP.


\section{Conclusions} \label{sec:S-Conclusions}

%
%
In this paper, we introduced and validated a new optimization method for model-data 
fitting, ROAM (Radial-basis-functions Optimization Approximation Method). Our primary 
motivation for this work has been to develop a novel approach for diagnosing the solar 
coronal magnetic field by combining a parameterized 3D magnetic field model 
with forward modeling of coronal polarization. From various tests applied to the synthetic 
test bed of a coronal magnetic flux rope, we showed that ROAM allows for fast, efficient, 
and accurate model-data fitting in a $d$-dimensional parameter space. These test cases 
further enabled us to analyze and specify a framework for an optimal application of 
ROAM.

%
%
Applying our method with forward modeling of IR coronal polarimetry, we demonstrated 
that ROAM can be exploited for converting coronal polarimetric measurements 
into magnetic field data. The use of our model-data fitting method therefore opens 
new perspectives for the development and exploitation of coronal polarimetric 
measurements such as the ones routinely performed by CoMP \citep{Tomczyk08} and 
future telescopes such as the Daniel K. Inoue Solar Telescope\footnote{\url{http://www.ifa.hawaii.edu/~schad/dlnirsp/}} 
and the Coronal Solar Magnetism Observatory (Tomczyk et al., submitted), 
but also for a wider range of coronal observations including, \eg UV \citep[see \eg][]{Fineschi01,Raouafi09} 
and radio polarimetry (\eg \citeauthor{White97}, \citeyear{White97}; \citeauthor{Gelfreikh04}, 
\citeyear{Gelfreikh04}; see also \citeauthor{Gibson16}, \citeyear{Gibson16}, for discussion 
of multiwavelength magnetometry).

%
%
Beyond the analysis of coronal polarimetric measurements, ROAM offers interesting 
perspectives for magnetic field reconstruction models. Most of the current 3D diagnostics 
of the coronal magnetic field of solar active regions (ARs) are derived from the analysis 
of magnetic field reconstruction models including, \eg force-free field extrapolations 
of the photospheric magnetic field 
\citep[see \eg][and references therein]{Alissandrakis81,Demoulin89,Wheatland00,Yan00,Wiegelmann04,Amari06,Malanushenko12}, 
{and} magneto-frictional methods 
\citep[see \eg][and references therein]{VanBallegooijen04,Valori05,Valori07,Jiang11,Inoue12,Titov14}. 
ROAM could, in principle, 
be used to perform model-data fitting with such reconstruction models that either already 
are  (\ie through the poloidal and axial flux for the {magneto-frictional methods with} 
flux rope insertion) or could be 
(\eg through the photospheric force-free parameter for both force-free field extrapolations 
and magneto-frictional methods {without flux rope insertion)} parameterized. 
{The extensive work performed} 
over the years in terms of forward modeling of various observables 
\citep[see \eg][and references therein]{Gibson16} would then allow for using several types 
of different observations to constrain the parameters of the magnetic field reconstruction 
models. ROAM therefore opens new perspectives for including coronal polarimetric 
measurements into magnetic field reconstructions and, more generally, for data-optimized 
reconstruction of the solar coronal magnetic field. Such perspectives will be tackled 
in the framework of the Data Optimized Coronal Field Model\footnote{\url{http://www.hao.ucar.edu/DOCFM/}} 
(DOCFM), a collaborative project that will make use of ROAM.

%
%
Finally, we wish to mention that ROAM is not limited to coronal magnetic field diagnostics 
and could be used for other optimization problems. The method will be of particular interest 
for model-data fitting for which a model evaluation (here, the evaluation of the model itself 
and/or the forward modeling of an observable if applicable) is computationally expensive.


\section*{Disclosure/Conflict-of-Interest Statement}

The authors declare that the research was conducted in the absence of any commercial or financial relationships that could be construed as a potential conflict of interest.


\section*{Author Contributions}

All the authors contributed to the building, writing, and editing of the content of the paper.


\section*{Acknowledgments}

We thank the two anonymous referees and Anna Malanushenko for a careful consideration 
of the manuscript and constructive comments. The National Center for Atmospheric Research 
is sponsored by the National Science Foundation.

\paragraph{Funding\textcolon} K.D. acknowledges funding from the Computational 
and Information Systems Laboratory and from the High Altitude Observatory, and 
along with S.E.G and Y.F. acknowledges support from the Air Force Office of Scientific 
Research under award FA9550-15-1-0030. 


\section*{Supplemental Data}
Supplementary Material should be uploaded separately on submission, if there are Supplementary Figures, please include the caption in the same file as the figure. LaTeX Supplementary Material templates can be found in the Frontiers LaTeX folder


\appendix
\section{Appendix: Solving the coefficients of the RBF decomposition} \label{app:A-Solving-coeffs}

The $a_j$ and $b_j$ coefficients of the RBF decomposition (\eq{Eq-RBF-Likelihood}) are determined 
from the constraint $\hat{\ell} \left( \xx_i \right) = \ell \left( \xx_i \right)$, which leads to the $n$ following 
equations
	\BE	\label{eq:Eq-L-nequations-0}
		 \ell (\xx_i) =  \sum^{n}_{j=1} a_j \varphi_{j} (\xx_i) + \sum^{N}_{j=1} b_j \psi_{j} (\xx_i)       \,.
	\EE 
where $N = \binom{p+d}{p}$; note that $N \ne n$. \eq{Eq-L-nequations-0} can then be re-written as
	\BE	\label{eq:Eq-L-nequations-1}
		 \ell (\xx_i) = 
				\begin{pmatrix} 
					\varphi_{1} (\xx_i) & \varphi_{2} (\xx_i) & \cdots & \varphi_{n} (\xx_i) 
				\end{pmatrix} 
				\begin{pmatrix}
					a_1 \\ 
					a_2 \\
					\vdots \\
					a_n
				\end{pmatrix}  
				+ 
				\begin{pmatrix} 
					\psi_{1} (\xx_i) & \psi_{2} (\xx_i) & \cdots & \psi_{N} (\xx_i) 
				\end{pmatrix} 
				\begin{pmatrix}
					b_1 \\
					b_{2} \\
					\vdots \\
					b_{N}
				\end{pmatrix}       \,,
	\EE
where \eq{Eq-L-nequations-1} is the $i$-th row of the matrix equation
	\BE	\label{eq:Eq-L-nequations-final}
		 \vec{L} = \Phi \vec{a} + \Psi \vec{b}    \,,
	\EE
with
	\BA	\label{eq:Eq-L-nequations-details}
		 \vec{L} =  \begin{pmatrix}
					\ell (\xx_1) \\
					\ell (\xx_2) \\
					\vdots \\
					\ell (\xx_n)
				\end{pmatrix}  \,, & 
		\vec{a} =  \begin{pmatrix}
					a_1 \\
					a_2 \\
					\vdots \\
					a_n
				\end{pmatrix} \,, & 
		\vec{b} =  \begin{pmatrix}
					b_0 \\
					b_1 \\
					\vdots \\
					b_{N}
				\end{pmatrix} \,,   \\
		\Phi =       \begin{pmatrix}
					\varphi_{1} (\xx_1) & \cdots & \varphi_{n} (\xx_1) \\
					\vdots &  & \vdots  \\
					\varphi_{1} (\xx_n) & \cdots & \varphi_{n} (\xx_n)
				\end{pmatrix} & 
				\,, &
		\Psi =           \begin{pmatrix}
					\psi_{1} (\xx_1) & \cdots & \psi_{N} (\xx_1) \\
					\vdots & & \vdots  \\
					\psi_{1} (\xx_n) & \cdots & \psi_{N} (\xx_n)
				\end{pmatrix} \,,
	\EA
where $\vec{L}$ and $\vec{a}$ are $n$-vectors, $\vec{b}$ is a $N$-vector, 
$\Phi$ is a $n \times n$ matrix, and $\Psi$ is a $n \times N$ matrix.

\eq{Eq-L-nequations-final} is expressed as the sum of an affine, $\Psi \vec{b}$, and 
non-affine, $\Phi \vec{a}$, term. The $\vec{a}$ and $\vec{b}$ coefficients can be 
solved by separating the affine and non-affine terms of \eq{Eq-L-nequations-final} 
through QR-factorization of $\Psi$. QR-factorization or QR-decomposition consists 
in the decomposition of a matrix $M$ into a product of an orthogonal matrix Q 
and an upper triangular matrix R, such that $M=QR$. The QR-factorization of 
our rectangular matrix $\Psi$ leads
	\BE   \label{eq:Eq-QR-general}
		\Psi = QR =  \begin{pmatrix}
					Q_1 & Q_2
				\end{pmatrix} 
				 \begin{pmatrix}
					R_1 \\
					0
				\end{pmatrix} = Q_1 R_1  \,,
	\EE
where $R_1$ is a $N \times N$ upper triangular matrix, $Q_1$ is $n \times N$, and $Q_2$ is 
$n \times (n - N)$ (which implies $n > N$). In QR-factorization with $n > N$, the columns of both 
$Q_1$ and $Q_2$ are orthonormal, which implies $Q^{T}_{1} Q^{}_{1} = I$ and 
$Q^{T}_{2} Q^{}_{2} = I$, where $T$ denotes the transpose. Note that $Q$ being an orthogonal 
matrix implies that each column of $Q_1$ is orthogonal to each column of $Q_2$, which means 
that $Q_2$ is in the orthogonal complement of $Q_1$ and $Q^{T}_{2} Q^{}_{1} = 0$ and 
$Q^{T}_{1} Q^{}_{2} = 0$. With that in mind, it is then possible to separate the affine and 
non-affine terms in \eq{Eq-L-nequations-final} by taking the $\vec{a}$ coefficients in the space 
associated with $Q_2$, such that $\vec{a} = Q_2 \vec{\gamma}$, where 
$\vec{\gamma}$ is a $N$-vector. Replacing this particular solution of $\vec{a}$ and multiplying 
\eq{Eq-L-nequations-final} by $Q^{T}_{2}$ leads to
	\BA   
		Q^{T}_{2} \vec{L} & = & Q^{T}_{2} \Phi Q^{}_{2} \vec{\gamma} + Q^{T}_{2} Q^{}_{1} R^{}_{1} \vec{b}  \nonumber   \\
		\label{eq:Eq-Gamma-tmp} 
		 & = &  Q^{T}_{2} \Phi Q^{}_{2} \vec{\gamma}    \,
	\EA
which, after some algebra, results in
	\BE   \label{eq:Eq-Gamma}
		\vec{\gamma} = (Q^{T}_{2} \Phi Q^{}_{2})^{-1} Q^{T}_{2} \vec{L}     \,.
	\EE
The $\vec{a}$ and $\vec{b}$ coefficients are then given by
	\BA   \label{eq:Eq-RBF-coefficients}
		\vec{a} & = & Q^{}_{2} (Q^{T}_{2} \Phi Q^{}_{2})^{-1} Q^{T}_{2} \vec{L}     \,,   \\
		\label{eq:Eq-Spline-coefficients}
		\vec{b} & = & R^{-1}_{1} Q^{T}_{1} \left( \vec{L} - \Phi \vec{a} \right)     \,.
	\EA

In all the results presented in this paper, the QR-factorization and the $\vec{a}$ and $\vec{b}$ 
coefficients were computed using the DGEQRF, DORGQR, and DGETRF routines of the LAPACK 
fortran library.


\section{Appendix: RBF decomposition with periodic components} \label{app:A-Periodic-Variables}

For all applications considered for our optimization method, the parameter space, $d$-space, 
always defines a set of points in the Euclidean space, $\mathbb{E}^{d'}$, where $d' \ge d$ 
(although the $d$-space may itself be non-Euclidean). The $d$-space belongs to $\mathbb{E}^d$ 
only when it does not possess any periodic parameter. Whenever they exist, periodic directions 
imply that the $d$-space is curved, which in turn means that the set of points of the $d$-space 
is actually defined in $\mathbb{E}^{d'}$ with ${d'} > d$. For instance, this is the case of the points 
defining the surface of a cylinder and/or a sphere. Although they define a two dimensional 
$d$-space, these points actually form a set of $\mathbb{R}^3$. Hence, a transformation from 
$\mathbb{R}^d$ to $\mathbb{R}^{d'}$ should be applied to the $\xx$ $d$-vectors if one wants 
the periodicities of the $d$-space to be included for the RBF decomposition and one must then 
substitute $d$ by $d'$ in \eqss{Eq-RBF-Likelihood}{Eq-RBF-2} (see \sect{S-Method}).

Let $\xx^T = \begin{pmatrix} \xx^{T}_{\mathrm{n-per.}},\xx^{T}_{\mathrm{per.}} \end{pmatrix}$ 
be a vector of the $d$-space such that $\xx_{\mathrm{n-per.}}$ and $\xx_{\mathrm{per.}}$ 
are the sets of non-periodic and periodic components of $\xx$. If all components of 
$\xx_{\mathrm{per.}}$ are independent, the coordinate transformation from $\mathbb{R}^d$ 
to $\mathbb{R}^{d'}$ is
	\BE	\label{eq:Eq-IPVT-1}
		\xx \mapsto  \xx' = 
		  		\begin{pmatrix}
					\xx_{\mathrm{n-per.}}    \\ 
					\vec{u}_1    \\
					\vec{u}_2
				\end{pmatrix}      \,, 
	\EE
where the $i$-th component of $\vec{u}_1$ and $\vec{u}_2$ is
	\BA	\label{eq:Eq-IPVT-2} 
		u_{i,1} & = & \cos \left( 2\pi \frac{x_{i,\mathrm{per.}}}{P_{i}} \right) \,,     \\
		u_{i,2} & = & \sin \left( 2\pi \frac{x_{i,\mathrm{per.}}}{P_{i}} \right)           \,,
	\EA
where $P_{i}$ is the period of $x_{i,\mathrm{per.}}$. For a $d$-space with independent periodic 
components, then $d' = d_{\mathrm{n-per.}} + 2d_{\mathrm{per.}}$ (where $d_{\mathrm{Y}}$ is 
the dimension of $\xx_{Y}$).

When some or all of the periodic components of the $d$-space are coupled, the coordinate 
transformation depends upon the relationship between the coupled periodic components. 
For instance in a 2D parameter space with $\xx = (x_1, x_2)$, if $x_1$ and $x_2$ are 
the longitude and co-latitude on a sphere ($x_1 \in [0 ; 2\pi]$ and $x_2 \in [0 ; \pi]$), then 
the coordinate transformation from $\mathbb{R}^d$ to $\mathbb{R}^{d'}$ is
	\BA	\label{eq:Eq-CPVT} 
		u_{1} & = & \cos \left( x_1 \right)  \sin \left( x_2 \right) \,,     \\
		u_{2} & = & \sin \left( x_1 \right) \sin \left( x_2 \right)   \,,     \\
		u_{3} & = & \cos \left( x_2 \right)           \,,
	\EA
leading to $d' = 3$.

  \begin{figure}
   \centerline{\includegraphics[width=0.75\textwidth,clip=]{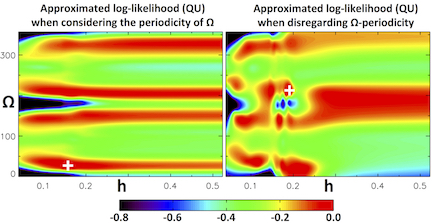}
              }
              \textbf{\refstepcounter{figure}\label{fig:Fig-2paras-LfS-period-vs-noperiod} Figure \arabic{figure}.}{ Comparison between approximated log-likelihood {functions} when taking the periodicity of $\Omega$ into account (left panel) and when disregarding it (right panel). The approximated log-likelihood {functions} are associated with the design of 2DN120 giving the best MLE (shown panel (c) of \fig{Fig-2paras-LfS-RB-dispersions}). The white ``+'' symbol indicates the position of the maximum log-likelihood. Notice the bumpy behavior of the approximated log-likelihood {function} when the periodicity of $\Omega$ is not accounted for and the MLE which is completely different from the MLE obtained when considering the periodicity.}
   \end{figure} 

Considering the periodicities is not mandatory for applying the RBF decomposition. 
However, not considering it may significantly alter the accuracy of the RBF decomposition, 
leading to a bumpy approximation and a bad MLE. 
Such an example is displayed in \fig{Fig-2paras-LfS-period-vs-noperiod} 
for one of the 2DN120 case (see details in \sect{S-Likelihood-multiple-maxima}). The bumps 
in the case that disregards the periodicity are likely to occur whenever the range of any 
periodic component of the $d$-space becomes strictly larger than half the associated 
period. That hypothesis is suggested by the 3DN301 and 3DN31ZOOM cases of \sect{S-Results} 
that give good log-likelihood {function} approximations while disregarding the periodicities 
of $\theta$, $\phi$, and $\Omega$ whose ranges were smaller than half the periods. 
The aforementioned hypothesis can also be hinted at from the following. Let $\xx_A$ and 
$\xx_B$ be two points in $\mathbb{R}^d$, $\xx_A$ is held fixed while $\xx_B$ is being 
moved away from $\xx_A$ along the periodic component only. In $\mathbb{R}^{d}$ 
where the periodicity is disregarded, the Euclidean distance between these two points 
always increases monotonically as $\xx_B$ is moved away. That is not the case in 
$\mathbb{R}^{d'}$ where periodicity is taken into account. Indeed, the Euclidean distance 
between these two points periodically goes through maxima and minima when the distance 
along the periodic component reaches half the period and the period, respectively (like 
what happens when one puts two points on a circle and moves one of the two points 
along the circle). As a consequence of its definition that only depends on Euclidean 
distances, an RBF in $\mathbb{R}^{d'}$ is expected to display a similar cyclic behavior 
along the periodic components. Therefore, the behavior of an RBF will be consistent, 
\ie monotonic, in both $\mathbb{R}^d$ and $\mathbb{R}^{d'}$ as long as the distance 
to that RBF along any periodic component remains smaller than half the corresponding 
period, and hence, similar results are expected for the RBF decomposition whether it is 
performed in $\mathbb{R}^d$ or $\mathbb{R}^{d'}$. This is not true anymore when the range 
of one or more periodic components becomes larger than half the associated period.

Finally, let us emphasize that the intuitive approach to deal with periodic components 
and that consists in computing the contribution of an RBF, $\varphi_A$, at $\xx$ by taking 
the minimum distance, $r_{\min}$, between $||\xx - \xx_A ||$ and $||\xx - \xx_A \pm \vec{P} ||$ 
(where $\vec{P}^T = \begin{pmatrix} 0,0,\cdots,P \end{pmatrix}$) is strongly {\it ill-advised}. 
Indeed, it would have the undesirable effect of creating an artificial discontinuity of the RBF 
at $||\xx_A \pm \vec{P} ||$ caused by the fact that our RBFs are not compact. Taking 
the minimum distance, $r_{\mathrm{min}}$, would only work with compact RBFs, provided 
that their radius of influence (\ie the radius below which they have a non-zero value) in 
the periodic direction be strictly smaller than the corresponding period.


\bibliographystyle{frontiersinSCNS}
      
\bibliography{ModelDataFitting}  

\begin{thebibliography}{54}
\providecommand{\natexlab}[1]{#1}
\expandafter\ifx\csname urlstyle\endcsname\relax
  \providecommand{\doi}[1]{doi:\discretionary{}{}{}#1}\else
  \providecommand{\doi}{doi:\discretionary{}{}{}\begingroup
  \urlstyle{rm}\Url}\fi
\providecommand{\selectlanguage}[1]{\relax}

\bibitem[{\textbf{{Alissandrakis}}(1981)}]{Alissandrakis81}
{Alissandrakis}, C.~E. (1981), {On the computation of constant alpha force-free
  magnetic field}, \emph{\aap}, 100, 197--200

\bibitem[{\textbf{{Amari} et~al.}(2006)\textbf{{Amari}, {Boulmezaoud}, and
  {Aly}}}]{Amari06}
{Amari}, T., {Boulmezaoud}, T.~Z., and {Aly}, J.~J. (2006), {Well posed
  reconstruction of the solar coronal magnetic field}, \emph{\aap}, 446,
  691--705, \doi{10.1051/0004-6361:20054076}

\bibitem[{\textbf{{Bird} et~al.}(1985)\textbf{{Bird}, {Volland}, {Howard},
  {Koomen}, {Michels}, {Sheeley} et~al.}}]{Bird85}
{Bird}, M.~K., {Volland}, H., {Howard}, R.~A., {Koomen}, M.~J., {Michels},
  D.~J., {Sheeley}, N.~R., Jr., et~al. (1985), {White-light and radio sounding
  observations of coronal transients}, \emph{\solphys}, 98, 341--368,
  \doi{10.1007/BF00152465}

\bibitem[{\textbf{{Bommier} and {Sahal-Brechot}}(1982)}]{Bommier82}
{Bommier}, V. and {Sahal-Brechot}, S. (1982), {The Hanle effect of the coronal
  L-alpha line of hydrogen - Theoretical investigation}, \emph{\solphys}, 78,
  157--178, \doi{10.1007/BF00151151}

\bibitem[{\textbf{Boser et~al.}(1992)\textbf{Boser, Guyon, and
  Vapnik}}]{Boser92}
Boser, B.~E., Guyon, I.~M., and Vapnik, V.~N. (1992), A training algorithm for
  optimal margin classifiers, in Proceedings of the Fifth Annual Workshop on
  Computational Learning Theory (ACM, New York, NY, USA), COLT '92, 144--152,
  \doi{10.1145/130385.130401}

\bibitem[{\textbf{{Broomhead} and {Lowe}}(1988)}]{Broomhead88}
{Broomhead}, D.~S. and {Lowe}, D. (1988), Multivariable functional
  interpolation and adaptive networks, \emph{Complex Systems}, 2, 321--355

\bibitem[{\textbf{{Buhmann}}(2003)}]{Buhmann03}
{Buhmann}, M.~D. (2003), Radial Basis Functions: Theory and Implementations,
  Cambridge Monographs on Applied and Computational Mathematics (Cambridge
  University Press)

\bibitem[{\textbf{{Casini} and {Judge}}(1999)}]{Casini99}
{Casini}, R. and {Judge}, P.~G. (1999), {Spectral Lines for Polarization
  Measurements of the Coronal Magnetic Field. II. Consistent Treatment of the
  Stokes Vector forMagnetic-Dipole Transitions}, \emph{\apj}, 522, 524--539,
  \doi{10.1086/307629}

\bibitem[{\textbf{{Demoulin} et~al.}(1989)\textbf{{Demoulin}, {Priest}, and
  {Anzer}}}]{Demoulin89}
{Demoulin}, P., {Priest}, E.~R., and {Anzer}, U. (1989), {A three-dimensional
  model for solar prominences}, \emph{\aap}, 221, 326--337

\bibitem[{\textbf{{Duchon}}(1977)}]{Duchon77}
{Duchon}, J. (1977), Constructive Theory of Functions of Several Variables:
  Proceedings of a Conference Held at Oberwolfach April 25 -- May 1, 1976
  (Springer Berlin Heidelberg, Berlin, Heidelberg), \doi{10.1007/BFb0086566}

\bibitem[{\textbf{{Fan}}(2012)}]{Fan12}
{Fan}, Y. (2012), {Thermal Signatures of Tether-cutting Reconnections in
  Pre-eruption Coronal Flux Ropes: Hot Central Voids in Coronal Cavities},
  \emph{\apj}, 758, 60, \doi{10.1088/0004-637X/758/1/60}

\bibitem[{\textbf{{Fasshauer}}(2007)}]{Fasshauer07}
{Fasshauer}, G.~E. (2007), Meshfree Approximation Methods with MATLAB, volume~6
  of \emph{Interdisciplinary Mathematical Sciences} (World Scientific
  Publishing Company)

\bibitem[{\textbf{{Fineschi}}(2001)}]{Fineschi01}
{Fineschi}, S. (2001), {Space-based Instrumentation for Magnetic Field Studies
  of Solar and Stellar Atmospheres}, in G.~{Mathys}, S.~K. {Solanki}, and D.~T.
  {Wickramasinghe}, eds., Magnetic Fields Across the Hertzsprung-Russell
  Diagram, volume 248 of \emph{Astronomical Society of the Pacific Conference
  Series}, volume 248 of \emph{Astronomical Society of the Pacific Conference
  Series}, 597

\bibitem[{\textbf{{Flyer} et~al.}(2016)\textbf{{Flyer}, {Barnett}, and
  {Wicker}}}]{Flyer16}
{Flyer}, N., {Barnett}, G.~A., and {Wicker}, L.~J. (2016), {Enhancing finite
  differences with radial basis functions: Experiments on the Navier-Stokes
  equations}, \emph{Journal of Computational Physics}, 316, 39--62,
  \doi{10.1016/j.jcp.2016.02.078}

\bibitem[{\textbf{{Flyer} and {Fornberg}}(2011)}]{Flyer11}
{Flyer}, N. and {Fornberg}, B. (2011), {Radial basis functions: Developments
  and applications to planetary scale flows}, \emph{Computers and Fluids}, 46,
  1, 23--32, \doi{http://dx.doi.org/10.1016/j.compfluid.2010.08.005}, 10th
  \{ICFD\} Conference Series on Numerical Methods for Fluid Dynamics (ICFD
  2010)

\bibitem[{\textbf{Fornberg and Flyer}(2015)}]{Fornberg15}
Fornberg, B. and Flyer, N. (2015), Solving pdes with radial basis functions,
  \emph{Acta Numerica}, 24, 215--258, \doi{10.1017/S0962492914000130}

\bibitem[{\textbf{{Gelfreikh}}(2004)}]{Gelfreikh04}
{Gelfreikh}, G.~B. (2004), {Coronal Magnetic Field Measurements Through
  Bremsstrahlung Emission}, in D.~E. {Gary} and C.~U. {Keller}, eds.,
  Astrophysics and Space Science Library, volume 314 of \emph{Astrophysics and
  Space Science Library}, volume 314 of \emph{Astrophysics and Space Science
  Library}, 115, \doi{10.1007/1-4020-2814-8_6}

\bibitem[{\textbf{Gibson et~al.}(2016)\textbf{Gibson, Kucera, White, Dove, Fan,
  Forland et~al.}}]{Gibson16}
Gibson, S., Kucera, T., White, S.~M., Dove, J., Fan, Y., Forland, B., et~al.
  (2016), {FORWARD: A toolsel for multiwavelength coronal magnetometry},
  \emph{Frontier in Astronomy and Space Sciences}, 3, 8,
  \doi{10.3389/fspas.2016.00008}

\bibitem[{\textbf{{Gibson} and {Charbonneau}}(1998)}]{Gibson98}
{Gibson}, S.~E. and {Charbonneau}, P. (1998), {Empirical modeling of the solar
  corona using genetic algorithms}, \emph{\jgr}, 103, 14511--14522,
  \doi{10.1029/98JA00676}

\bibitem[{\textbf{{Hale}}(1908)}]{Hale08}
{Hale}, G.~E. (1908), {On the Probable Existence of a Magnetic Field in
  Sun-Spots}, \emph{\apj}, 28, 315, \doi{10.1086/141602}

\bibitem[{\textbf{{Hanle}}(1924)}]{Hanle24}
{Hanle}, W. (1924), {{\"U}ber magnetische Beeinflussung der Polarisation der
  Resonanzfluoreszenz}, \emph{Zeitschrift fur Physik}, 30, 93--105,
  \doi{10.1007/BF01331827}

\bibitem[{\textbf{{Harder} and {Desmarais}}(1972)}]{Harder72}
{Harder}, R.~L. and {Desmarais}, R.~N. (1972), Interpolation using surface
  splines., \emph{Journal of Aircraft}, 9, 2, 189--191, \doi{10.2514/3.44330}

\bibitem[{\textbf{{Iman} et~al.}(1981)\textbf{{Iman}, {Helton}, and
  {Campbell}}}]{Iman81}
{Iman}, R.~L., {Helton}, J.~C., and {Campbell}, J.~E. (1981), An approach to
  sensitivity analysis of computer models, part 1. introduction, input variable
  selection and preliminary variable assessment, \emph{Journal of Quality
  Technology}, 13, 3, 174--183

\bibitem[{\textbf{{Inoue} et~al.}(2012)\textbf{{Inoue}, {Magara}, {Watari}, and
  {Choe}}}]{Inoue12}
{Inoue}, S., {Magara}, T., {Watari}, S., and {Choe}, G.~S. (2012), {Nonlinear
  Force-free Modeling of a Three-dimensional Sigmoid Observed on the Sun},
  \emph{\apj}, 747, 65, \doi{10.1088/0004-637X/747/1/65}

\bibitem[{\textbf{{Jiang} et~al.}(2011)\textbf{{Jiang}, {Feng}, {Fan}, and
  {Xiang}}}]{Jiang11}
{Jiang}, C., {Feng}, X., {Fan}, Y., and {Xiang}, C. (2011), {Reconstruction of
  the Coronal Magnetic Field Using the CESE-MHD Method}, \emph{\apj}, 727, 101,
  \doi{10.1088/0004-637X/727/2/101}

\bibitem[{\textbf{Jones et~al.}(1998)\textbf{Jones, Schonlau, and
  Welch}}]{Jones98}
Jones, D.~R., Schonlau, M., and Welch, W.~J. (1998), Efficient global
  optimization of expensive black-box functions, \emph{Journal of Global
  Optimization}, 13, 4, 455--492, \doi{10.1023/A:1008306431147}

\bibitem[{\textbf{{Judge}}(2007)}]{Judge07}
{Judge}, P.~G. (2007), {Spectral Lines for Polarization Measurements of the
  Coronal Magnetic Field. V. Information Content of Magnetic Dipole Lines},
  \emph{\apj}, 662, 677--690, \doi{10.1086/515433}

\bibitem[{\textbf{{Judge} et~al.}(2006)\textbf{{Judge}, {Low}, and
  {Casini}}}]{Judge06}
{Judge}, P.~G., {Low}, B.~C., and {Casini}, R. (2006), {Spectral Lines for
  Polarization Measurements of the Coronal Magnetic Field. IV. Stokes Signals
  in Current-carrying Fields}, \emph{\apj}, 651, 1229--1237,
  \doi{10.1086/507982}

\bibitem[{\textbf{{Kramar} et~al.}(2014)\textbf{{Kramar}, {Airapetian},
  {Miki{\'c}}, and {Davila}}}]{Kramar14}
{Kramar}, M., {Airapetian}, V., {Miki{\'c}}, Z., and {Davila}, J. (2014), {3D
  Coronal Density Reconstruction and Retrieving the Magnetic Field Structure
  during Solar Minimum}, \emph{\solphys}, 289, 2927--2944,
  \doi{10.1007/s11207-014-0525-7}

\bibitem[{\textbf{{Lam}}(1983)}]{Lam83}
{Lam}, N.~S.~N. (1983), {Spatial interpolation methods: A review}, \emph{The
  American Cartographer}, 10, 2, 129--149

\bibitem[{\textbf{{Lin} et~al.}(2004)\textbf{{Lin}, {Kuhn}, and
  {Coulter}}}]{Lin04}
{Lin}, H., {Kuhn}, J.~R., and {Coulter}, R. (2004), {Coronal Magnetic Field
  Measurements}, \emph{\apjl}, 613, L177--L180, \doi{10.1086/425217}

\bibitem[{\textbf{{L{\'o}pez Ariste}}(2015)}]{LopezAriste15}
{L{\'o}pez Ariste}, A. (2015), {Magnetometry of Prominences}, in J.-C. {Vial}
  and O.~{Engvold}, eds., Solar Prominences, volume 415 of \emph{Astrophysics
  and Space Science Library}, volume 415 of \emph{Astrophysics and Space
  Science Library}, 179, \doi{10.1007/978-3-319-10416-4_8}

\bibitem[{\textbf{Louis and Rawlins}(1992)}]{Louis92}
Louis, S.~J. and Rawlins, G. J.~E. (1992), Predicting convergence time for
  genetic algorithms, in Foundations of Genetic Algorithms 2 (Morgan Kaufmann),
  141--161

\bibitem[{\textbf{{Madych} and {Nelson}}(1990)}]{Madych90}
{Madych}, W.~R. and {Nelson}, S.~A. (1990), Polyharmonic cardinal splines,
  \emph{Journal of Approximation Theory}, 60, 2, 141 -- 156,
  \doi{http://dx.doi.org/10.1016/0021-9045(90)90079-6}

\bibitem[{\textbf{{Malanushenko} et~al.}(2012)\textbf{{Malanushenko},
  {Schrijver}, {DeRosa}, {Wheatland}, and {Gilchrist}}}]{Malanushenko12}
{Malanushenko}, A., {Schrijver}, C.~J., {DeRosa}, M.~L., {Wheatland}, M.~S.,
  and {Gilchrist}, S.~A. (2012), {Guiding Nonlinear Force-free Modeling Using
  Coronal Observations: First Results Using a Quasi-Grad-Rubin Scheme},
  \emph{\apj}, 756, 153, \doi{10.1088/0004-637X/756/2/153}

\bibitem[{\textbf{{McKay} et~al.}(1979)\textbf{{McKay}, {Beckman}, and
  {Conover}}}]{McKay79}
{McKay}, M.~D., {Beckman}, R.~J., and {Conover}, W.~J. (1979), A comparison of
  three methods for selecting values of input variables in the analysis of
  output from a computer code, \emph{Technometrics}, 21, 2, 239--245

\bibitem[{\textbf{{Nychka} et~al.}(2015)\textbf{{Nychka}, {Bandyopadhyay},
  {Hammerling}, {Lindgren}, and {Sain}}}]{Nychka15}
{Nychka}, D., {Bandyopadhyay}, S., {Hammerling}, D., {Lindgren}, F., and
  {Sain}, S. (2015), A multiresolution gaussian process model for the analysis
  of large spatial datasets, \emph{Journal of Computational and Graphical
  Statistics}, 24, 2, 579--599, \doi{10.1080/10618600.2014.914946}

\bibitem[{\textbf{{Plowman}}(2014)}]{Plowman14}
{Plowman}, J. (2014), {Single-point Inversion of the Coronal Magnetic Field},
  \emph{\apj}, 792, 23, \doi{10.1088/0004-637X/792/1/23}

\bibitem[{\textbf{Powell}(1977)}]{Powell77}
Powell, M.~J.~D. (1977), Restart procedures for the conjugate gradient method,
  \emph{Mathematical Programming}, 12, 1, 241--254, \doi{10.1007/BF01593790}

\bibitem[{\textbf{Powell}(1993)}]{Powell93}
Powell, M.~J.~D. (1993), Some algorithms for thin plate spline interpolation to
  functions of two variables, \emph{Cambridge University Dept. of Applied
  Mathematics and Theoretical Physics technical report}

\bibitem[{\textbf{{Rachmeler} et~al.}(2012)\textbf{{Rachmeler}, {Casini}, and
  {Gibson}}}]{Rachmeler12}
{Rachmeler}, L.~A., {Casini}, R., and {Gibson}, S.~E. (2012), {Interpreting
  Coronal Polarization Observations}, in T.~R. {Rimmele}, A.~{Tritschler},
  F.~{W{\"o}ger}, M.~{Collados Vera}, H.~{Socas-Navarro}, R.~{Schlichenmaier},
  M.~{Carlsson}, T.~{Berger}, A.~{Cadavid}, P.~R. {Gilbert}, P.~R. {Goode}, and
  M.~{Kn{\"o}lker}, eds., Second ATST-EAST Meeting: Magnetic Fields from the
  Photosphere to the Corona., volume 463 of \emph{Astronomical Society of the
  Pacific Conference Series}, volume 463 of \emph{Astronomical Society of the
  Pacific Conference Series}, 227

\bibitem[{\textbf{{Rachmeler} et~al.}(2013)\textbf{{Rachmeler}, {Gibson},
  {Dove}, {DeVore}, and {Fan}}}]{Rachmeler13}
{Rachmeler}, L.~A., {Gibson}, S.~E., {Dove}, J.~B., {DeVore}, C.~R., and {Fan},
  Y. (2013), {Polarimetric Properties of Flux Ropes and Sheared Arcades in
  Coronal Prominence Cavities}, \emph{\solphys}, 288, 617--636,
  \doi{10.1007/s11207-013-0325-5}

\bibitem[{\textbf{{Raouafi} et~al.}(2009)\textbf{{Raouafi}, {Solanki}, and
  {Wiegelmann}}}]{Raouafi09}
{Raouafi}, N.-E., {Solanki}, S.~K., and {Wiegelmann}, T. (2009), {Hanle Effect
  Diagnostics of the Coronal Magnetic Field: A Test Using Realistic Magnetic
  Field Configurations}, in S.~V. {Berdyugina}, K.~N. {Nagendra}, and
  R.~{Ramelli}, eds., Solar Polarization 5: In Honor of Jan Stenflo, volume 405
  of \emph{Astronomical Society of the Pacific Conference Series}, volume 405
  of \emph{Astronomical Society of the Pacific Conference Series}, 429

\bibitem[{\textbf{{Sahal-Brechot} et~al.}(1977)\textbf{{Sahal-Brechot},
  {Bommier}, and {Leroy}}}]{SahalBrechot77}
{Sahal-Brechot}, S., {Bommier}, V., and {Leroy}, J.~L. (1977), {The Hanle
  effect and the determination of magnetic fields in solar prominences},
  \emph{\aap}, 59, 223--231

\bibitem[{\textbf{{Stenflo}}(2015)}]{Stenflo15}
{Stenflo}, J.~O. (2015), {History of Solar Magnetic Fields Since George Ellery
  Hale}, \emph{\ssr}, \doi{10.1007/s11214-015-0198-z}

\bibitem[{\textbf{{Titov} et~al.}(2014)\textbf{{Titov}, {T{\"o}r{\"o}k},
  {Mikic}, and {Linker}}}]{Titov14}
{Titov}, V.~S., {T{\"o}r{\"o}k}, T., {Mikic}, Z., and {Linker}, J.~A. (2014),
  {A Method for Embedding Circular Force-free Flux Ropes in Potential Magnetic
  Fields}, \emph{\apj}, 790, 163, \doi{10.1088/0004-637X/790/2/163}

\bibitem[{\textbf{{Tomczyk} et~al.}(2008)\textbf{{Tomczyk}, {Card}, {Darnell},
  {Elmore}, {Lull}, {Nelson} et~al.}}]{Tomczyk08}
{Tomczyk}, S., {Card}, G.~L., {Darnell}, T., {Elmore}, D.~F., {Lull}, R.,
  {Nelson}, P.~G., et~al. (2008), {An Instrument to Measure Coronal Emission
  Line Polarization}, \emph{\solphys}, 247, 411--428,
  \doi{10.1007/s11207-007-9103-6}

\bibitem[{\textbf{{Valori} et~al.}(2007)\textbf{{Valori}, {Kliem}, and
  {Fuhrmann}}}]{Valori07}
{Valori}, G., {Kliem}, B., and {Fuhrmann}, M. (2007), {Magnetofrictional
  Extrapolations of Low and Lou's Force-Free Equilibria}, \emph{\solphys}, 245,
  263--285, \doi{10.1007/s11207-007-9046-y}

\bibitem[{\textbf{{Valori} et~al.}(2005)\textbf{{Valori}, {Kliem}, and
  {Keppens}}}]{Valori05}
{Valori}, G., {Kliem}, B., and {Keppens}, R. (2005), {Extrapolation of a
  nonlinear force-free field containing a highly twisted magnetic loop},
  \emph{\aap}, 433, 335--347, \doi{10.1051/0004-6361:20042008}

\bibitem[{\textbf{{van Ballegooijen}}(2004)}]{VanBallegooijen04}
{van Ballegooijen}, A.~A. (2004), {Observations and Modeling of a Filament on
  the Sun}, \emph{\apj}, 612, 519--529, \doi{10.1086/422512}

\bibitem[{\textbf{{Wheatland} et~al.}(2000)\textbf{{Wheatland}, {Sturrock}, and
  {Roumeliotis}}}]{Wheatland00}
{Wheatland}, M.~S., {Sturrock}, P.~A., and {Roumeliotis}, G. (2000), {An
  Optimization Approach to Reconstructing Force-free Fields}, \emph{\apj}, 540,
  1150--1155, \doi{10.1086/309355}

\bibitem[{\textbf{{White} and {Kundu}}(1997)}]{White97}
{White}, S.~M. and {Kundu}, M.~R. (1997), {Radio Observations of Gyroresonance
  Emission from Coronal Magnetic Fields}, \emph{\solphys}, 174, 31--52,
  \doi{10.1023/A:1004975528106}

\bibitem[{\textbf{{Wiegelmann}}(2004)}]{Wiegelmann04}
{Wiegelmann}, T. (2004), {Optimization code with weighting function for the
  reconstruction of coronal magnetic fields}, \emph{\solphys}, 219, 87--108,
  \doi{10.1023/B:SOLA.0000021799.39465.36}

\bibitem[{\textbf{{Yan} and {Sakurai}}(2000)}]{Yan00}
{Yan}, Y. and {Sakurai}, T. (2000), {New Boundary Integral Equation
  Representation for Finite Energy Force-Free Magnetic Fields in Open Space
  above the Sun}, \emph{\solphys}, 195, 89--109, \doi{10.1023/A:1005248128673}

\end{thebibliography}

\IfFileExists{\jobname.bbl}{} {\typeout{}
\typeout{****************************************************}
\typeout{****************************************************}
\typeout{** Please run "bibtex \jobname" to obtain} \typeout{**
the bibliography and then re-run LaTeX} \typeout{** twice to fix
the references !}
\typeout{****************************************************}
\typeout{****************************************************}
\typeout{}}








\end{document}